\documentclass[twocolumn]{aastex63}
\usepackage[utf8]{inputenc}
\usepackage{amsmath}
\usepackage{amsfonts}
\usepackage{amssymb}
\usepackage{graphicx}
\usepackage{hyperref}
\usepackage[caption=false]{subfig}
\usepackage{tabularx}
\usepackage{array}
\usepackage{xcolor}
\usepackage{listings}
\usepackage{comment}
\usepackage{bm}
\usepackage{slashed}
\usepackage{algorithm}
\usepackage[noend]{algpseudocode}

\usepackage{siunitx}
\usepackage{multirow}
\usepackage{booktabs}
\usepackage[para]{threeparttable}
\usepackage[normalem]{ulem} 
\usepackage{textcomp} 

\newcommand*{\inc}{\Delta}

\newcommand*{\dk}{\textsc{DarkKROME }}
\newcommand*{\dkns}{\textsc{DarkKROME}}
\newcommand*{\krome}{\textsc{KROME }}
\newcommand*{\kromens}{\textsc{KROME}}

\newcommand{\code}[1]{\texttt{#1}}
\newcommand*{\ddm}{dissipative dark matter }
\newcommand*{\ddmns}{dissipative dark matter}

\newcommand*{\ra}[1]{r_{\alpha}^{#1}}
\newcommand*{\rc}[1]{r_{m}^{#1}}
\newcommand*{\rx}[1]{r_{M}^{#1}}
\newcommand*{\racx}[3]{\ra{#1}\, \rc{#2}\, \rx{#3}}
\newcommand*{\rac}[2]{\ra{#1}\, \rc{#2}}
\newcommand*{\rax}[2]{\ra{#1}\, \rx{#2}}

\newcommand*{\rde}{r_{\inc E}}

\newcommand*{\KE}{\text{K.E.}}
\newcommand*{\ed}{\text{e}_{\rm D}}
\newcommand*{\hd}{\text{H}_{\rm D}}
\newcommand*{\hdt}{\text{H}_{\rm D,2}}

\DeclareSIUnit \atomicunit{a.u.}
\DeclareSIUnit \erg{erg}
\DeclareSIUnit \rydberg{Ry}
\DeclareSIUnit \year{yr}
\DeclareSIUnit \epccm{\erg\per\centi\meter\cubed\per\second}
\DeclareSIUnit \clight{\text{\ensuremath{c}}}
\newcommand{\pcc}{\per\centi\meter\cubed}

\begin{document}

	\title{Molecular Chemistry for Dark Matter III: \dk}
	\date{\today}
	
	\author[0000-0002-0378-5195]{Michael Ryan}
	\email{mzr55@psu.edu}
	\affiliation{Institute for Gravitation and the Cosmos, The Pennsylvania State University, University Park, PA 16802, USA}
	\affiliation{Department of Physics, The Pennsylvania State University, University Park, PA, 16802, USA}
	
	\author[0000-0002-6498-6812]{Sarah Shandera}
	\email{ses47@psu.edu}
	\affiliation{Institute for Gravitation and the Cosmos, The Pennsylvania State University, University Park, PA 16802, USA}
	\affiliation{Department of Physics, The Pennsylvania State University, University Park, PA, 16802, USA}
	
	\author[0000-0002-8677-1038]{James Gurian}
	\email{jhg5248@psu.edu}
	\affiliation{Institute for Gravitation and the Cosmos, The Pennsylvania State University, University Park, PA 16802, USA}
	\affiliation{Department of Astronomy and Astrophysics, The Pennsylvania State University, University Park, PA, 16802, USA}
	
	\author[0000-0002-8434-979X]{Donghui Jeong}
	\email{djeong@psu.edu}
	\affiliation{Institute for Gravitation and the Cosmos, The Pennsylvania State University, University Park, PA 16802, USA}
	\affiliation{Department of Astronomy and Astrophysics, The Pennsylvania State University, University Park, PA, 16802, USA}
	\affiliation{School of Physics, Korea Institute for Advanced Study (KIAS), 85 Hoegiro, Dongdaemun-gu, Seoul, 02455, Republic of Korea}

\begin{abstract}
Dark matter that is dissipative may cool sufficiently to form compact objects, including black holes. Determining the abundance and mass spectrum of those objects requires an accurate model of the chemistry relevant for the cooling of the dark matter gas. Here we introduce a chemistry tool for dark matter, DarkKROME, an extension of the KROME software package. DarkKROME is designed to include all atomic and molecular processes relevant for dark matter with two unequal-mass fundamental fermions, interacting via a massless-photon mediated $U(1)$ force. We use DarkKROME to perform one-zone collapse simulations and 
study the evolution of temperature-density phase diagrams for various dark-sector parameters. DarkKROME is publicly available at \url{https://bitbucket.org/mtryan83/darkkrome}.
\end{abstract}
\keywords{cosmology: theory -- dark matter -- molecular processes}
	
\section{Introduction} \label{sec:intro}
	
	Around 84 percent of all matter in the universe appears to be non-baryonic \citep{Planck2016}; its composition is still unknown. Current constraints on the particle nature of this dark matter are inferred from cosmological data~\citep{Planck:2018vyg}, from data on the gravitationally bound structures in the universe on scales ranging from about 10 Mpc - 100 pc \citep{Sofue:2000jx,McConnachie2012,Strigari:2012acq,Collins2021}, and via dark-matter particle detection experiments on Earth \citep{Schumann:2019eaa,Lin:2019uvt,Agrawal:2021dbo,Billard:2021uyg}. Gravitational wave observations from compact object mergers can also constrain dark matter models, an approach that is increasingly constructive given the recent success of current ground-based observatories \citep{PhysRevD.40.3221,Abbott:2005pf,Abbott:2007xi,Kouvaris:2010jy,deLavallaz:2010wp,Bramante:2014zca,Bramante:2015dfa,Bird:2016dcv,Sasaki:2016jop,Bramante:2017ulk,Kouvaris:2018wnh,LIGOScientific:2018zbj,Gresham:2018rqo,Diego:2019rzc,Authors:2019qbw,Gow:2019pok,DeLuca:2020qqa,Singh2020,Nitz:2020bdb,Nitz:2021mzz, Nitz:2021vqh}. To date, only gravitational interactions between dark and visible matter have been observed. 
	
	The traditional dark-matter model of the WIMP (weakly interacting massive particle), where dark matter has a very simple particle content, has an attractive simplicity. But, dark matter need not be so minimal. Some observations \citep{Bullock:2017xww,2020Univ....6..107D,Cyr-Racine:2021alc} and some theoretical considerations \citep{Zurek:2013wia,Petraki:2013wwa,Arkani-Hamed:2016rle,Chacko:2018vss} suggest that more complex physics may be at work. If dark matter has a richer particle content, it likely has internal chemistry. And, if dark matter has chemistry, it may be able to dissipate sufficient kinetic energy through scattering and atomic- or molecular-like transitions to cool and form compact objects, including black holes \citep{Cline2014,DAmico2017,Shandera2018,Latif2019}. In that case, fully modeling the evolution of dark-matter structures (and the baryonic structures that trace, albeit biased, the dark structure) requires new numerical tools to evolve the dark matter gas, including all relevant scattering and chemical processes. The modeling in turn will enable data on the abundance and mass spectrum of black holes from gravitational wave observatories to be used to constrain the particle properties of dark matter.

	A particular model of dissipative dark matter that is complex yet calculable is the ``atomic" dark matter scenario \citep{goldberg_new_1986,Ackerman2009,Feng2009,Kaplan2010,Kaplan2011,CyrRacine2013,cyr-racine/etal:2014,Fan2013,Cline2014,2015PhRvD..91b3512F,2016JCAP...07..013F,2015JCAP...09..057R,Boddy2016,Agrawal2017,Ghalsasi:2017jna}. Here dark matter consists of a heavy fermion with mass $M$ and a light fermion with mass $m$. These particles are oppositely charged by a $U(1)$ force of strength $\alpha_D$, mediated by a dark photon, $\gamma_D$, that allows the particles to form atoms and molecules, $\hd$ and $\hdt$, nearly analogous to atomic and molecular hydrogen. When $M\gg m$, dark-molecular processes can be obtained by a simple re-scaling of Standard Model processes \citep{Ryan2021}. There are no weak or strong force analogs in this model, so no quarks, neutrons, muons, etc. While this simplicity may seem ad hoc, it allows the model to be treated in precise numerical detail, providing an important benchmark scenario to calibrate phenomenological treatments applicable to dissipative scenarios more generally. 
	
	Cosmologically, an important additional parameter is the ratio of the dark-photon background temperature to the standard-photon background temperature, $\xi=T_{\gamma,D} / T_{CMB}$. We allow data considerations to drive the choice of temperature (that is, no input assumption that the dark matter and Standard Model were thermalized at any point) and use $\xi\ll0.4$ to be consistent with constraints on additional light degrees of freedom and the lack of observed dark acoustic oscillations \citep{cyr-racine/etal:2014}. We also assume the existence of a dark matter/anti-matter asymmetry, such that matter dominates, and the net dark $U(1)$ charge of the universe is $\approx0$ \citep{Kaplan2010}.
	
	In two companion papers we have derived the molecular physics of atomic dark matter \citep{Ryan2021}, and computed the cosmological abundances of the atoms and molecules for a wide range of parameter values \citep{Gurian2021}. Here, we present an application of those works: an extension of the software package \kromens ~\citep{Grassi2014}, used to implement chemical and thermal evolution in astrophysical and cosmological simulations, to evolve the chemical network of atomic dark matter. This
    extension, \dkns, is a tool that can help bridge the gap between the semi-analytic models of dissipative dark matter, simulations of structure on cluster and galaxy scales, and future gravitational-wave observations.

	This article is organized as follows. In Section \ref{sec:krome_and_dk}, we introduce the \dk extension, explaining how it implements the dissipative dark-matter model.
	In Section \ref{sec:simulation_model} we show how the results of a simple one-zone collapse simulation can vary depending on the masses and coupling strength of the dark matter. We conclude in Section \ref{sec:conclusion}.
	
\section{Introduction to \dk} \label{sec:krome_and_dk}
	\kromens{}\footnote{The \krome software package and documentation are available at \url{kromepackage.org}.}\citep{Grassi2014} is a tool to evaluate the thermal and chemical evolution of astrophysical gas clouds, assuming Standard Model particle content. It is used to generate a library of function calls which implement the user-specified reactions at the current gas composition, temperature, and timestep. This library can then be included in simulations of the gravitational dynamics of the gas. \krome provides several simple example simulations, including one-zone spherical cloud collapse and a one
    dimensional spherical shock, or it can be embedded into more complex simulations \citep{Suazo2019,Latif2019,Capelo2018,Prieto2015}. \krome leaves all dark-matter related physics, including any possible chemistry, to the exterior simulation. \dk extends \krome by enabling it to include dissipative dark reactions and thermal processes in the atomic dark matter model (i.e. the dark chemistry), without changing the overall library structure. \dk is publicly available under the GNU GPLv3 license and can be found at \url{https://bitbucket.org/mtryan83/darkkrome}.

	The \krome software package provides a \code{Python} pre-processor that takes as input a list of chemical reactions and heating and cooling processes, and produces \code{FORTRAN} subroutines. These are used by the \code{krome} function call to solve the set of (usually) sparse, stiff ordinary differential equations for the time evolution of particle number densities $n_i$ and the gas temperature $T$, 
	\begin{align}
		\frac{d n_i}{dt} = &\sum_{j \in F_i}\left(k_{j} \prod_{r \in R_j} n_{r(j)} \right) - \sum_{j \in D_i}\left(k_{j} \prod_{r \in R_j'} n_{r(j)}\right) \\
		\frac{dT}{dt}=&(\gamma-1)\frac{\Gamma(T,\bar{n})-\Lambda(T,\bar{n})}{k_B \sum_i n_i}\,.
		\label{eq:odes}
	\end{align} 
	In the first equation, the change in number density of species $i$ is determined by all the formation reactions $F_i$, with rates $k_{j}$ and reactants $R_j$, and all the destruction reactions $D_i$, with reactants $R_j'$. In the second equation, the change in temperature depends on the adiabatic index, $\gamma$, the collective heating, $\Gamma$, and cooling, $\Lambda$, (both in $\si{\erg\per\centi\meter\cubed\per\second}$) in addition to the particle number densities $\bar{n}=\{n_i\}$ and
    Boltzmann Constant, $k_B$~\citep{Grassi2014}. 
	
	\dk extends the \krome framework to support a new class of chemical species, designated with a $Q$ (i.e. $\hd$ would be \code{QH}), representing a second sector with a chemistry entirely  decoupled from that of the Standard Model, i.e. the dark sector.\footnote{We have chosen $Q$ such that there is no possible overlap with a Standard Model element name.} The sub-atomic particles included are a dark photon, QG, and the two fundamental fermions in the atomic dark matter model, with masses
    \code{qp\_mass} ($M$ earlier), and \code{qe\_mass} ($m$). These masses, along with the parameters \code{Dalpha} ($\alpha_D$) and \code{xi} ($\xi$), are additional inputs to \dk with default values equal to their Standard Model analogs. As an example usage, the masses and \code{Dalpha} are used in the reaction rates found in the chemical network, \code{react\_dark}, and all four parameters are part of several new thermal process rates. We have not built any dark antimatter parameters into
    \dkns, although they could be included by the user in the normal \krome fashion. 
	The list of dark matter model parameters can be found in the first section of Table \ref{tab:parameters}.
	\begin{table}[h]
		\centering
		\begin{threeparttable}
			\begin{tabular*}{\columnwidth}{l l l}
				\toprule
				\textbf{Parameter} & & \textbf{Description} \\
				\midrule
				\code{qe\_mass} & & dark electron mass: $m$\\
				\code{qp\_mass} & & dark proton mass: $M$\\
				\code{Dalpha} & & dark fine structure constant: $\alpha_D$\\
				\multirow{2}{*}{\code{xi}} & & ratio of dark photon background \\
				& & temperature to CMB: $T_{\gamma,D}/T_{CMB}$\\
				\midrule
				\code{DARKATOM} & & Re-scaled atomic cooling \\
				\code{ADARKATOM} & & Analytic atomic cooling/heating\\
				\code{DARKMOL} & & Re-scaled molecular cooling/heating \\
				\bottomrule
			\end{tabular*}
			\caption{Table of primary constants and rates added in \dk. The first section lists the main parameters of the dissipative dark matter model used, and the second lists the sets of cooling/heating processes included. The main constants are defined as possible additional parameters in the chemistry network files.}
			\label{tab:parameters}
		\end{threeparttable}
		
	\end{table}
	
	As appropriate for the atomic-dark-matter model, we have duplicated or extended the \krome subroutines to account for these new species, including reaction number and charge balancing, computing mean molecular weight and adiabatic index, etc. \krome provides significant additional machinery to compute grain physics, cosmic ray chemistry, advanced photochemistry, and other optional features, but we have not implemented corresponding dark versions beyond including a $Q$-photon, dark CMB flux (\code{darkCMB}), and the reaction and heating rates for $\hd$ photoionization.
	
	We have included three sets of thermal processes: re-scaled dark atomic cooling (with \code{-cooling=DARKATOM}), analytic dark atomic cooling and heating (with \code{-cooling=ADARKATOM} and \code{-heating=ADARKATOM}), and dark molecular cooling and heating (with \code{-cooling=DARKMOL} and \code{-heating=DARKMOL}). The re-scaled rates are obtained by extracting the dominant parametric dependence on $m$, $M$, and $\alpha$ for each process. Then, given a temperature-dependent rate $\Lambda_{\rm SM} (T)$ for a Standard Model process, the re-scaled rate appropriate for the corresponding process in the dark matter is given by 
	\begin{equation}
		\Lambda_{\rm DM} (T)=r_{\Lambda}\Lambda_{\rm SM} (\tilde{T})\,, 
	\end{equation}
	where $r_{\Lambda}$ is a dimensionless product of ratios of dark-to-standard-model parameters, and $\tilde{T}$ is the temperature re-scaled by ratio of dark to Standard Model (predominantly atomic) energy scales. Detailed expressions for the atomic processes, with Standard-Model processes from \citet{Cen1992} as used in \kromens, can be found in Appendix \ref{sec:atomic_cooling}. The analytic rates are those derived in \citet{Rosenberg2017}, with some implementation details in Appendix \ref{sec:atomic_cooling}. The analytic rates suffer from increased computational complexity and run time. Note that since the \code{-cooling=DARKATOM} and \code{-cooling=ADARKATOM} options include the same atomic processes, they are mutually exclusive.
	
\begin{figure}[htbp]
		\centering
		\includegraphics[width=\linewidth]{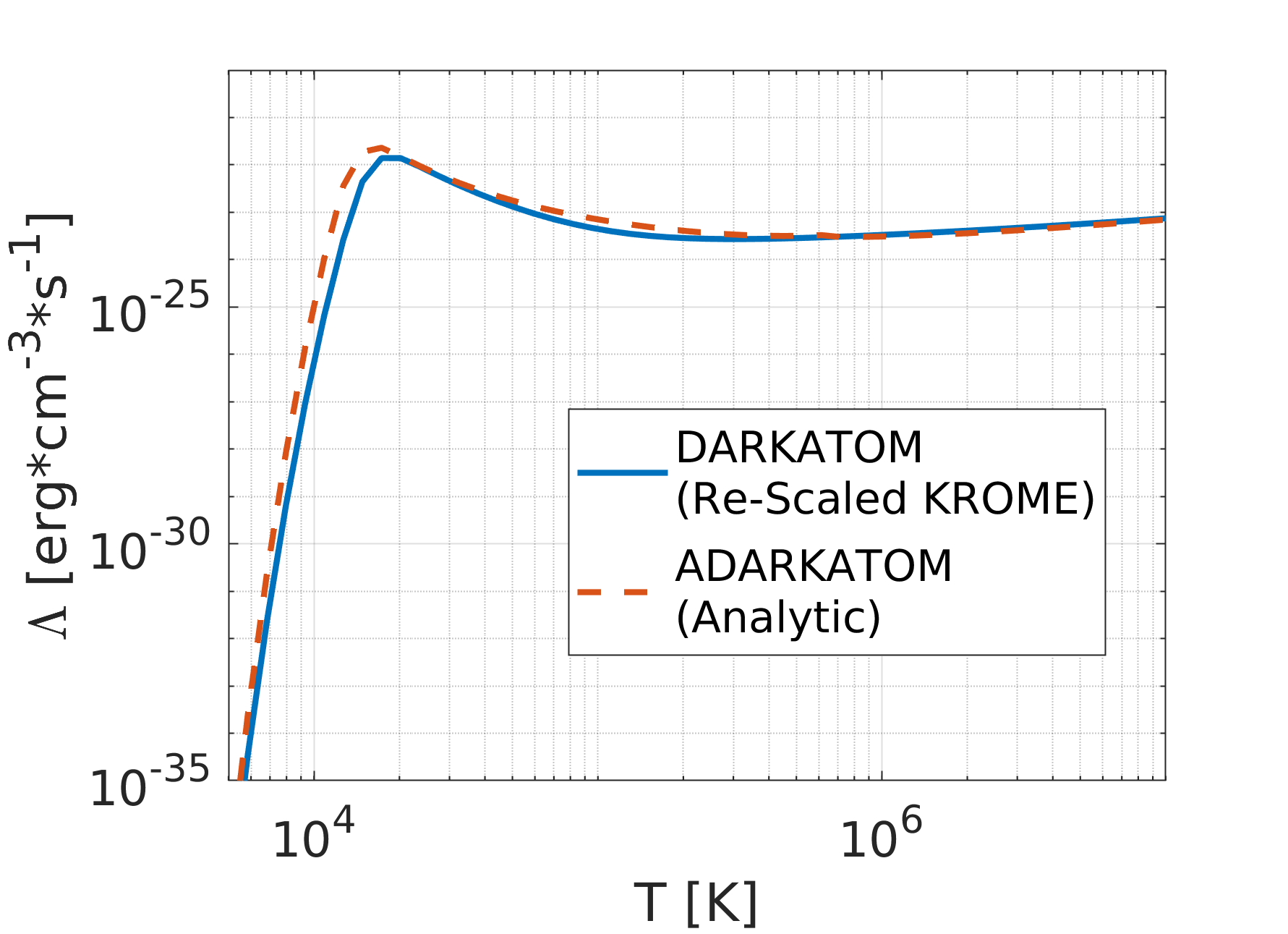}
        \caption{Comparison of the \code{DARKATOM} (re-scaled) and \code{ADARKATOM}(analytic) atomic cooling functions with Standard Model parameter values, $\xi=0.01$, and assuming chemical equilibrium at a particle density of \SI{1}{\pcc}. The difference between the re-scaled and analytic collisional excitation rates dominates the variation between the two rates, with more detail in Appendix \ref{sec:atomic_cooling}. }
		\label{fig:KROMEvsDARKnet}
	\end{figure}

	The dark atomic cooling rates, both re-scaled and analytic, include contributions from inverse Compton scattering, bremsstrahlung, collisional ionization, collisional excitation, and recombination, and are the dark analogs of the \code{-cooling=ATOMIC} plus \code{-cooling=COMPTON} and \code{-cooling=FF} rates in \krome. The analytic dark atomic heating rate only includes photoionization and is analogous to \code{-heating=PHOTO}. Figure \ref{fig:KROMEvsDARKnet} shows the slight difference in the net cooling rate, $\Lambda$, between the re-scaled and analytic approaches. The difference is predominantly due to the expressions for the collisional excitation cooling rate, and is of a similar level to the difference in Standard Model rates from \citet{Cen1992} and those found in other literature (e.g., \cite{Abel1997}). More detailed comparison is provided in Appendix \ref{sec:atomic_cooling}. 
	
	The dark molecular cooling rate includes contributions from dark molecular hydrogen ($\hdt$) rovibrational cooling and endoergic reactions, analogous to \code{-cooling=H2} plus \code{-cooling=CHEM}, while the heating rate includes several exoergic reactions, similar to \code{-heating=CHEM}. These re-scaled dark-matter rates were derived in \cite{Ryan2021}. We give additional implementation details in Appendix \ref{sec:dmc}. Since the rates are computed by re-scaling the pre-factors and temperature dependence of the built-in \krome $\text{H}_{\rm 2}$ rovibrational cooling rates from \cite{Glover2015}, \cite{Glover2008} and chemical thermal rates from \cite{Omukai2000}, the rates are identical to the \krome rates when Standard Model parameter values are used.
	
	As with the original \kromens, chemical reaction networks are included as external files, specified during the call to the \code{darkkrome} \textsc{Python} pre-processor. We have included an example chemical reaction network, \code{react\_dark}, that contains a minimal chemical network for primordial cloud collapse, as well as some ancillary variable definitions. More details of the \code{react\_dark} file can be found in Appendix \ref{sec:reactions}, but there are two important limitations as compared to common primordial chemical networks like the examples provided by \krome (e.g. \code{react\_primordial*}), or used in early universe literature (e.g. \citet{Galli1998,Glover2015}). First, the network does not include several subdominant $\hdt$ and $\hdt^+$ destruction reactions, such as $\hdt+\ed\rightarrow \hd+\hd^-$, which reduces the accuracy at temperatures approaching the $\hdt$ dissociation temperature and above. Second, the network only includes the minimal set of 3-body reactions from \citet{Ryan2021} and does not include any  reactions that involve $\text{H}_{D,3}$, which become important at high ($n_{\rm tot}>\SI{e8}{\pcc}(m/\SI{511}{\kilo\electronvolt})^3(\alpha/137^{-1})^3$) densities \citep{Glover2012,Gurian2021}. We leave the addition of these reactions to future work. 
	
\section{One-Zone Collapse} \label{sec:simulation_model}
	
	To verify \dk, we explored a simple density evolution model: a one-zone, uniform density cloud-collapse. First, we demonstrate that \dk can reproduce the results from running the one-zone collapse model built into \krome (\code{-test=CollapseZ}) with zero metallicity (i.e. only hydrogen and helium in the Standard Model), where the density, $\rho$, follows free-fall or adiabatic evolution, 
	\begin{equation}
		\frac{d\rho}{dt}=\frac{\rho}{t_{ff}},
		\label{eq:ffevol}
	\end{equation}
	where the free-fall time $t_{ff} = \sqrt{(3\pi) / (32 G \rho)}$, and $G$ is the gravitational constant.  The thermal processes include \code{DARKATOM/ADARKATOM} and \code{DARKMOL} heating and cooling, and compressional (adiabatic) heating, defined as 
	\begin{equation}
		\Gamma_{\rm compress} = \frac{\sum_i n_i k_B T}{t_{ff}}.
		\label{eq:compress}
	\end{equation}
	We end the simulation before the optically thick regime and so ignore the continuum cooling included in the CollapseZ test. For the CollapseZ comparison, we use the initial chemistry parameters including total particle density and species abundances specified in \citet{Grassi2014}.
	
	Second, we demonstrate that \dk can reproduce results found in \ddm literature \citep{DAmico2017}, which previously used the one-zone collapse model provided with \krome to model a ``mirror" dark sector, where particle content and parameter values are the same as in the Standard Model, except for $\xi$. In that example, the density evolution follows Equation \ref{eq:ffevol} unless the sound-crossing time $t_s=R/c_s$ is shorter than the free-fall time, wherein it follows isobaric evolution 
	\begin{equation}
		\frac{d\rho}{dt}=-\frac{T_0 \rho_0}{T^2}\frac{dT}{dt}.
		\label{eq:isoevol}
	\end{equation}
    In these equations, $c_s$ is the speed of sound in the gas, $R$ is the cloud radius, and $T_0$ is the initial temperature. The compressional heating term is turned off during isobaric evolution. Essentially, this allows for a more accurate evolution of the cloud wherein collapse only occurs if sound waves cannot traverse the cloud faster than a free-fall time. 
	
	The initial conditions of these simulations are determined by the cosmological parameters and the primary dark parameters, which set the initial chemistry variables including total particle density and abundance. We use the values for the cosmological parameters from \cite{Planck2016}, and set $\xi=0.01$. We further define the fraction of dissipative dark matter (out of total dark matter), $\epsilon$. Given $\xi$ and $\epsilon$, the primordial species abundances at the time of
    structure formation provide the initial species abundances in the cloud, where the primordial species abundance is either taken directly from \cite{DAmico2017} for comparison with that work or computed using the results from \cite{Gurian2021}. Lastly, we need to specify the initial temperature, provided as an input to the simulation. The required cosmological parameters are listed in part 1 of Table \ref{tab:sim_parameters}, with the initial chemistry parameters in part 2. 
	
    For further comparison, we show the results of varying the dark parameters, including $\xi$, and $\epsilon$. Doing so introduces additional categories of behaviors and indicates the potential for a wide range in dark, collapsed, halo mass scales.

	\begin{table}[htb!]
		\centering
		\begin{threeparttable}
			\begin{tabular*}{\columnwidth}{l l l}
				\toprule
				\textbf{Parameter} & & \textbf{Description} \\ 
				\midrule
				$h$ & & reduced Hubble constant \\
				$\delta_V$ & & virialization overdensity \\
				$\Omega_{M}$  &  & cosmological matter density\\
				$\Omega_{DM}$ &  & cosmological dark matter density\\
				\multirow{2}{*}{$\epsilon$} & & dissipative dark matter to all\\
				& & dark matter fraction: $\Omega_{DDM}/\Omega_{DM}$\\
				$z_s$ & & redshift at structure formation\\
				\midrule
                $n_{\rm tot}$ & & total particle number density of DDM \\
				\multirow{2}{*}{$n_0(Q)= x_0(Q)\;n_{\rm tot} $} & & initial number density of \\                                                 & & dark species $Q$ \\
				$T_0$ & & initial temperatue \\
				\bottomrule
			\end{tabular*} 
			\caption{Table of required cosmological parameters and initial chemical parameters required for the one-zone collapse simulations. Cosmological values combined with the dark parameters are used to compute the initial chemical parameters if not specified directly. Cosmological values are from \cite{Planck2016}.}
			\label{tab:sim_parameters}
		\end{threeparttable}
		
	\end{table}

\subsection{Verification Results} \label{sec:results}
	
	To check the reproduction, we consider the temperature evolution as a function of total particle density, or $n_{\rm tot}$ vs $T$. This is common practice in the literature (see e.g. \citet{Glover2008,Grassi2014,Yoshida2006,DAmico2017}). The Standard Model one-zone behavior has a direct analog in the radial temperature profile in full 3D hydrodynamical simulations, as seen in \citet{Yoshida2006} and \citet{Latif2019}, and consists of initial virialization heating, followed by efficient $\text{H}_2$ rovibrational cooling until density saturation, followed by rapid molecularization due to three-body reactions. 
	
\begin{figure}[htbp!]
		\centering
		\includegraphics[width=\linewidth]{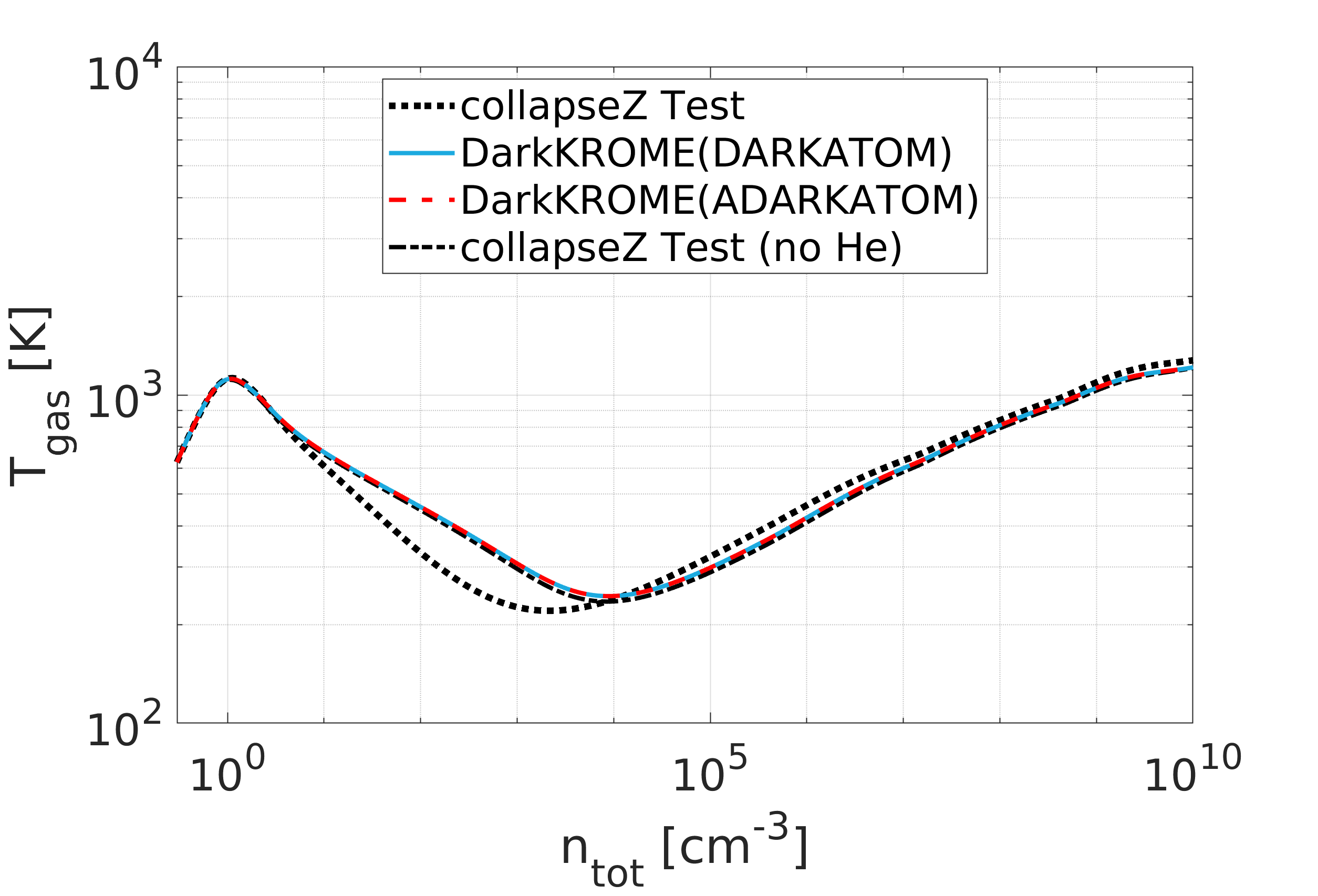} 
		\caption{Comparison of \krome and \dk primordial one-zone collapse. Note that the \dk curves exhibit less cooling than the \code{CollapseZ} test, as they lack collisional interactions with elements heavier than hydrogen, but otherwise demonstrate the expected qualitative behavior. The figure also demonstrates that there is minimal discernible difference between the different dark atomic cooling options. 
			\label{fig:onezonekromevdk}}
	\end{figure}

	Figure \ref{fig:onezonekromevdk} demonstrates that \dk can reproduce the results of running the primordial one-zone collapse simulation, \code{-test=CollapseZ}, provided in \kromens, using the initial fractional abundances of $x_{\ed}=x_{\hd^+}=\num{e-4}$, $x_{\hdt}=\num{e-6}$, and $x_{\hd}\approx 1$ and starting gas temperature $T_{\rm gas}=450 \text{K}$. \dk very nearly matches the results of \kromens, running with zero metallicity and the amount of helium set to zero. If helium is
    included, there are additional collisional channels that cool the gas at low densities. \dk does not currently contain the exothermic $H_{D,3}$ reactions that significantly heat the gas at densities above $\SI{e8}{\per\centi\meter\cubed}(m/\SI{511}{\kilo\electronvolt})^3(\alpha/137^{-1})^3$.
	
\begin{figure*}[hbtp!]
		\centering
		\begin{tabular}{cc}
			\includegraphics[width=0.47\textwidth]{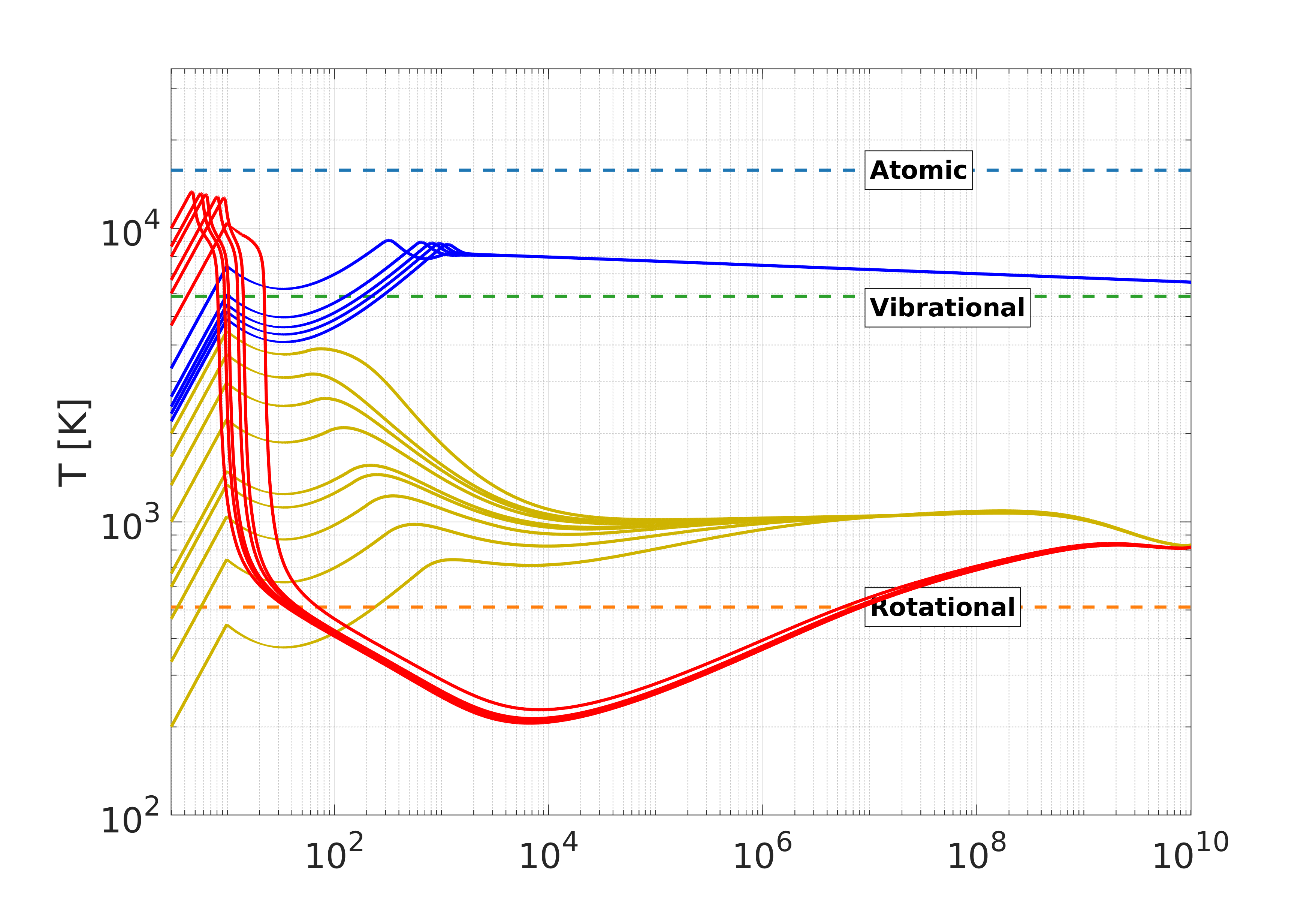} & 
			\includegraphics[width=0.47\textwidth]{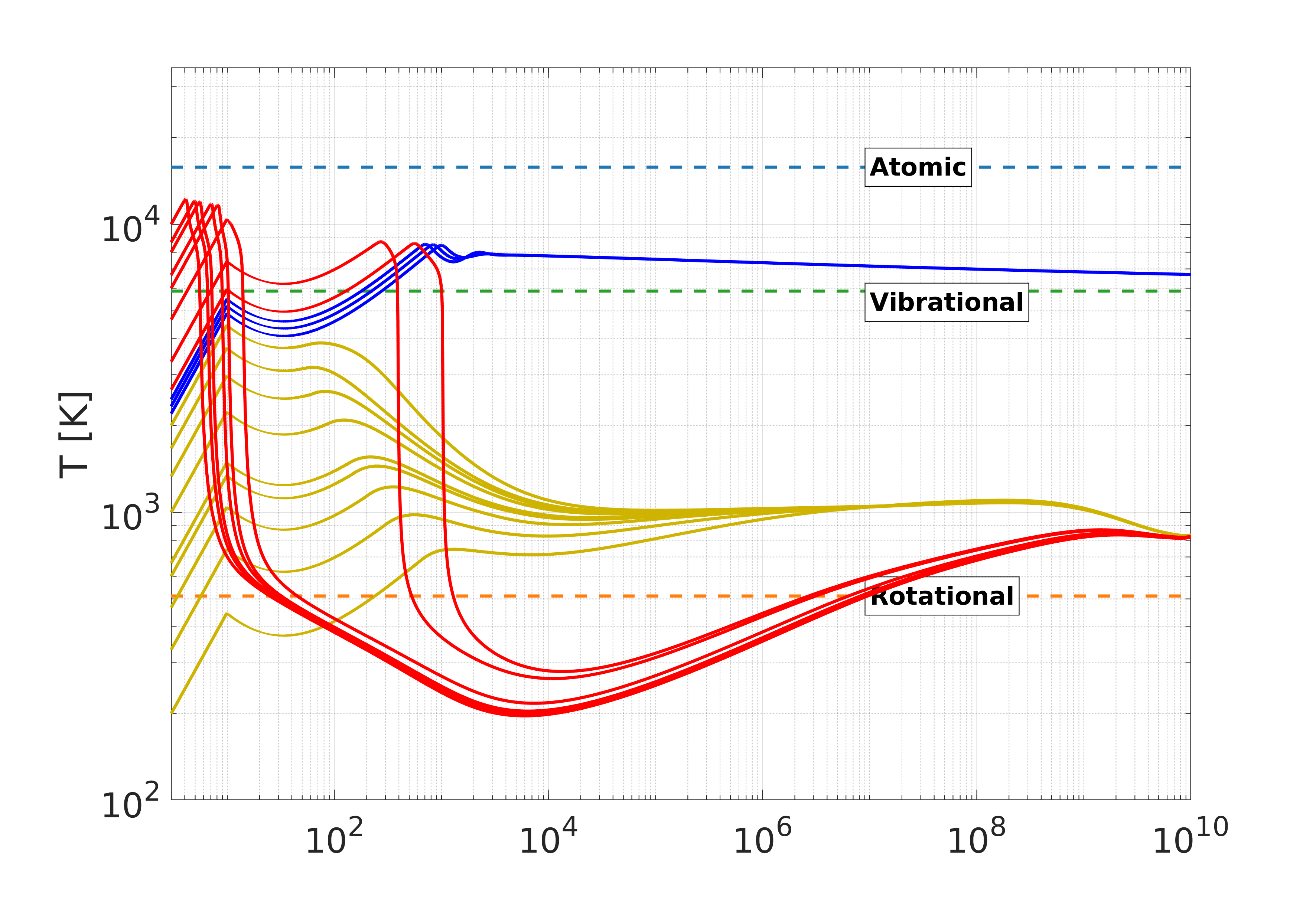} \\ 
			\includegraphics[width=0.47\textwidth]{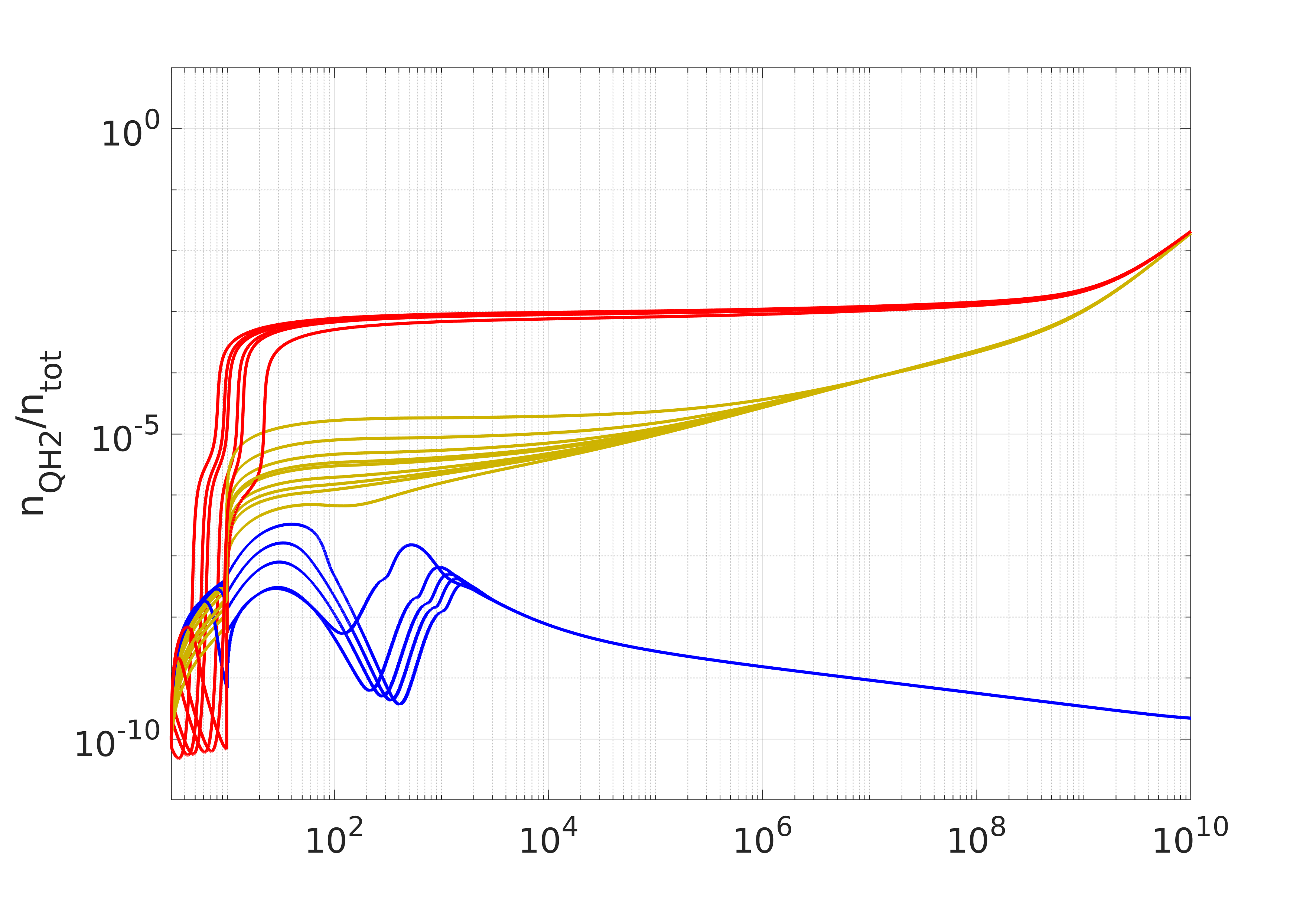} & 
			\includegraphics[width=0.47\textwidth]{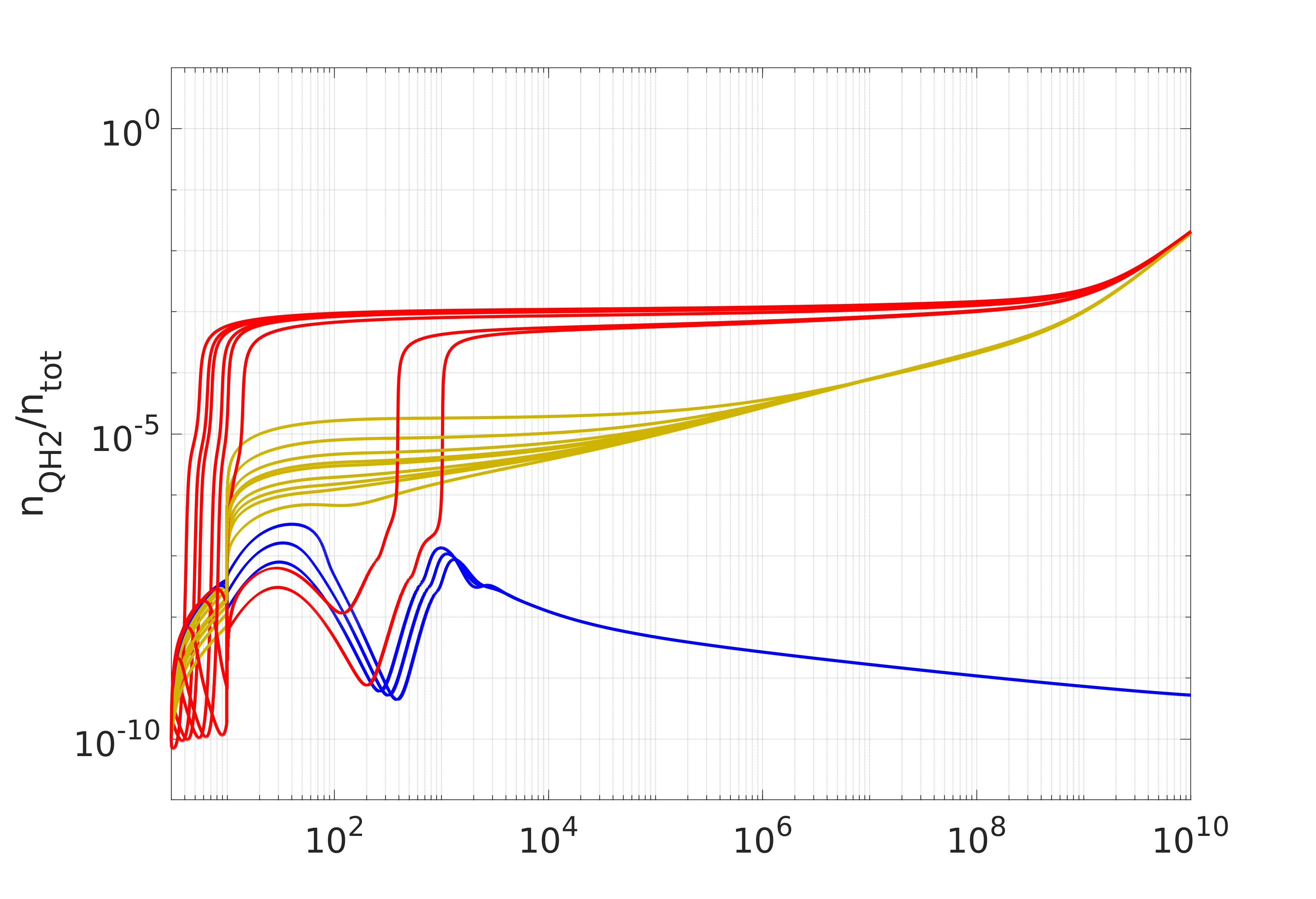} \\ 
			\includegraphics[width=0.47\textwidth]{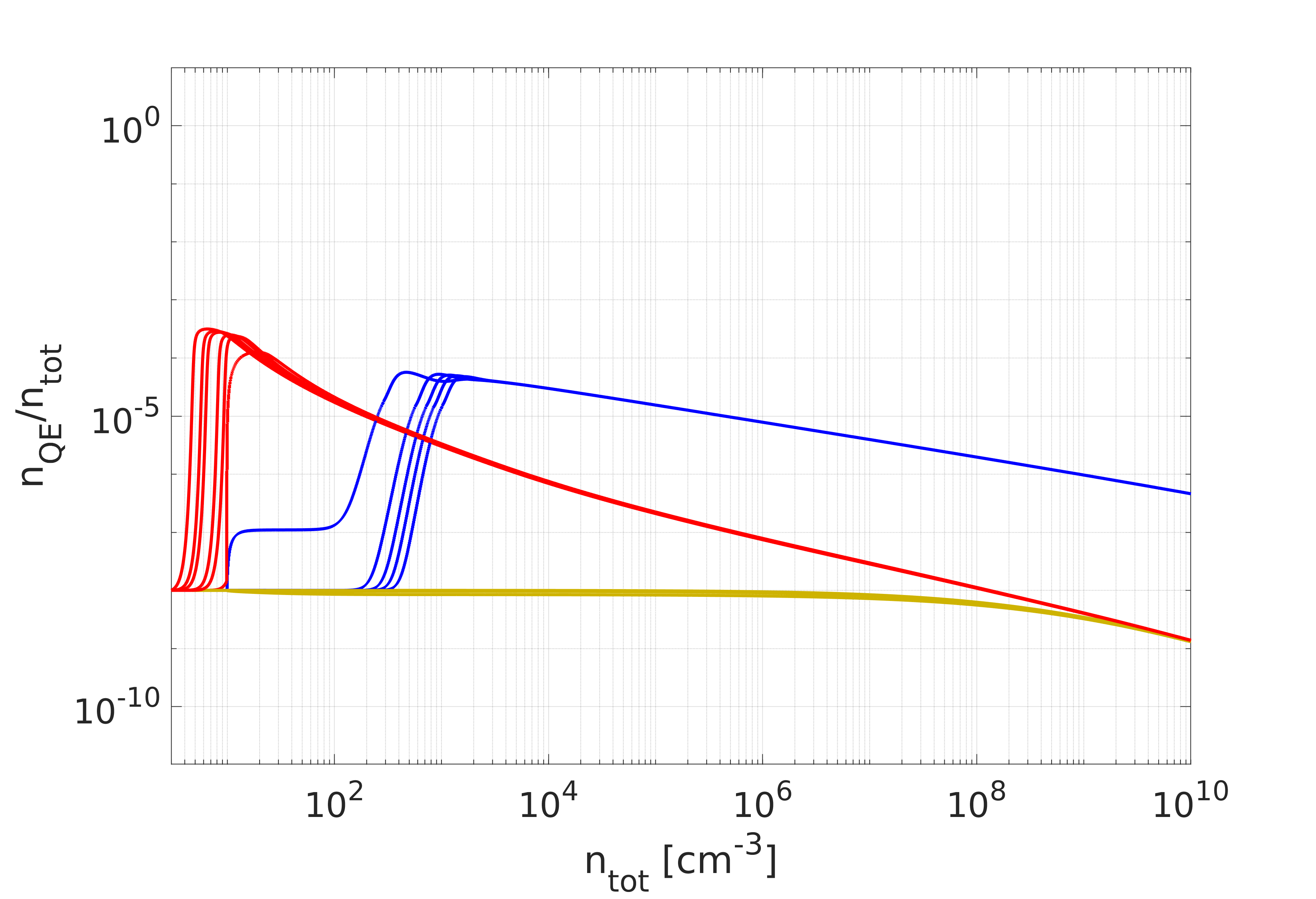} & 
			\includegraphics[width=0.47\textwidth]{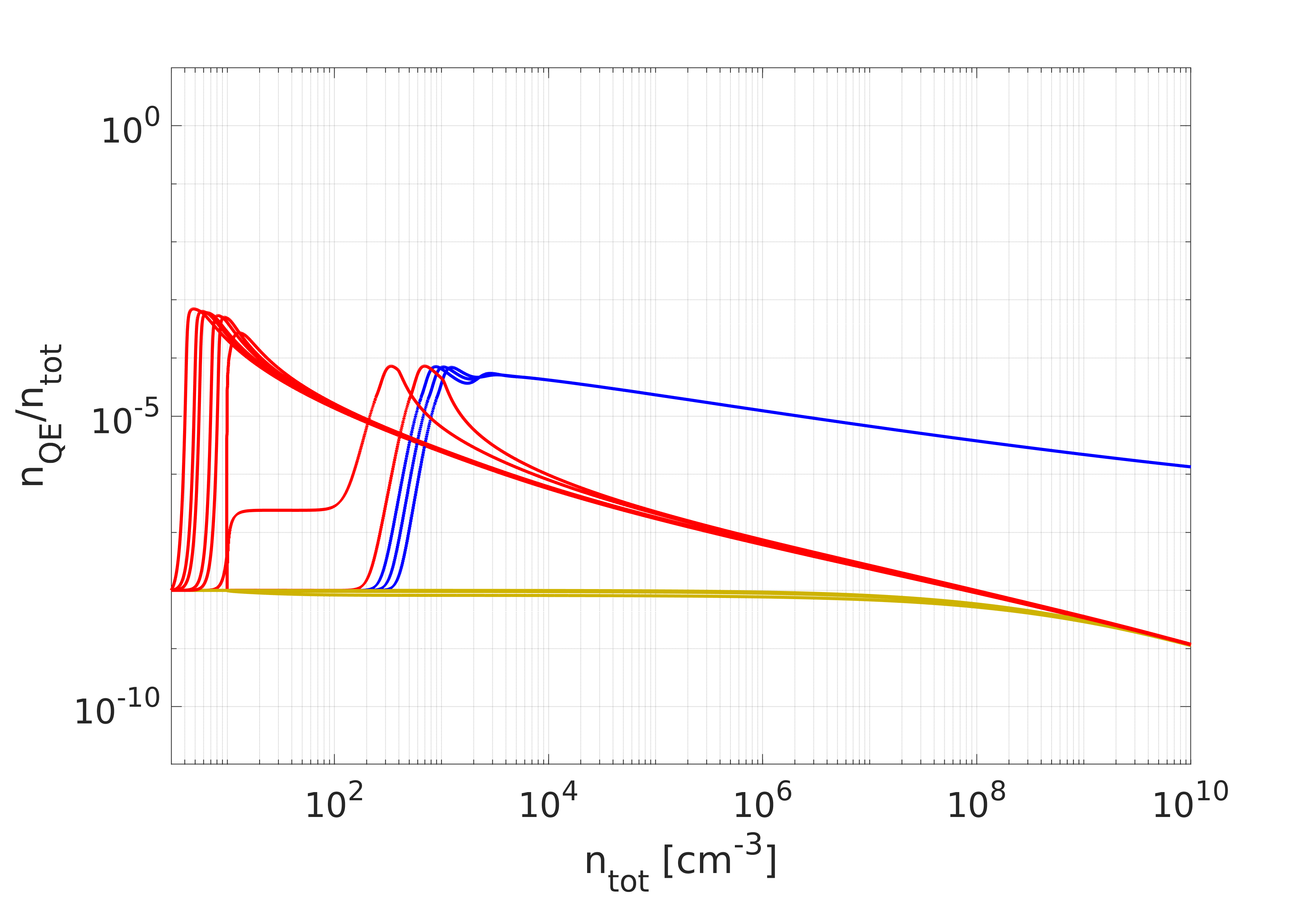} \\ 
		\end{tabular}
		
		\caption{
            Comparison between mirror dark matter results (equivalent to \citet{DAmico2017}) and atomic dark matter for the $\xi=0.01$ case. We have assumed $x_{\hdt}=\num{e-10}$, which is substantially lower than the actual cosmological abundance, for purposes of comparison with \citet{DAmico2017}. In the left column we run the one-zone collapse simulation for a ``mirror" dark sector, using the standard \kromens ~processes for a dissipative dark matter fraction of $\epsilon=0.18$. In the right column we use the reduced chemical network of atomic dark matter, described in Section \ref{sec:reactions}, and the \code{ADARKATOMIC} and \code{DARKMOL} cooling options (Table \ref{tab:sim_parameters}). The first row displays the temperature evolution as a function of total particle density and the middle and bottom row display the dark molecule $\hdt$(\code{QH2}) and dark electron $\ed$ (\code{QE}) abundances. Each line shows the evolution with different initial temperature $T_0$ and line colors follow the definitions in \citet{DAmico2017}, with red denoting efficient gas cooling, yellow denoting quasi-isothermal collapse in the range $(500-900)\si{\kelvin}$, and blue denoting quasi-isothermal collapse at $9000 \si{\kelvin}$. Also plotted are various temperature thresholds: the lowest $\hdt$ rotational and vibrational energy transitions, and the low-temperature peak of atomic collisional excitation cooling. Not shown is the $\hdt$ dissociation temperature, at approximately \SI{5e4}{\kelvin}. }
		\label{fig:damico_comp}
	\end{figure*}

	In Figure \ref{fig:damico_comp}, we reproduce the $\xi=0.01$ results of \cite{DAmico2017}, with the built-in Standard Model \krome chemical and thermal processes on the left and our added atomic dark matter model chemistry and thermal processes on the right. For both simulations we have used a starting total particle number density of $n_{\rm tot}=\SI{2.6}{\pcc}$, corresponding to $z=40$ and $\Omega_{\rm ADM}=\Omega_{\rm b}$ ($\epsilon=\num{0.18}$) and initial abundances of
    $x_{\ed}=x_{\hd^+}=\num{e-8}$, $x_{\hdt}=\num{e-10}$, and $x_{\hd}\approx 1$, following \cite{DAmico2017}.  While the atomic dark matter model does not contain dark helium, the amount of mirror helium is negligible at $\xi=0.01$ and does not factor into the comparison \citep{Berezhiani2001}. Each line corresponds to a different initial virial temperature, in the range $T_0$=\SIrange{300}{15000}{\kelvin}. Both simulations show three categories of trajectories: efficient molecular cooling (red lines), low-temperature quasi-isothermal collapse due to low free ionization (yellow lines), and high-temperature quasi-isothermal collapse due to delayed $\hdt$ formation (blue lines). The dark chemical network used here contains the relevant reactions to reproduce the main behavior categories, and in general matches the trajectories.
    There are, however, some differences shown in Figure~\ref{fig:damico_comp}, because we did not include some of the reactions used in the more complete network from \citet{DAmico2017}. In particular, there are fewer $\hdt^+$ destruction channels, leading to slightly higher $\hdt$ formation and cooling rates in some cases. This difference explains the discrepancy where the highest (lowest) high-temperature quasi-isothermal Standard Model trajectories in the left column of the figure are converted into efficient molecular cooling (low-temperature quasi-isothermal) dark trajectories, in the right column.
	
\subsection{Parameter Exploration Results} \label{sec:parameter_exploration}
	
\begin{figure*}[htbp!]
		\begin{tabular}{cc}
			\includegraphics[width=0.47\textwidth]{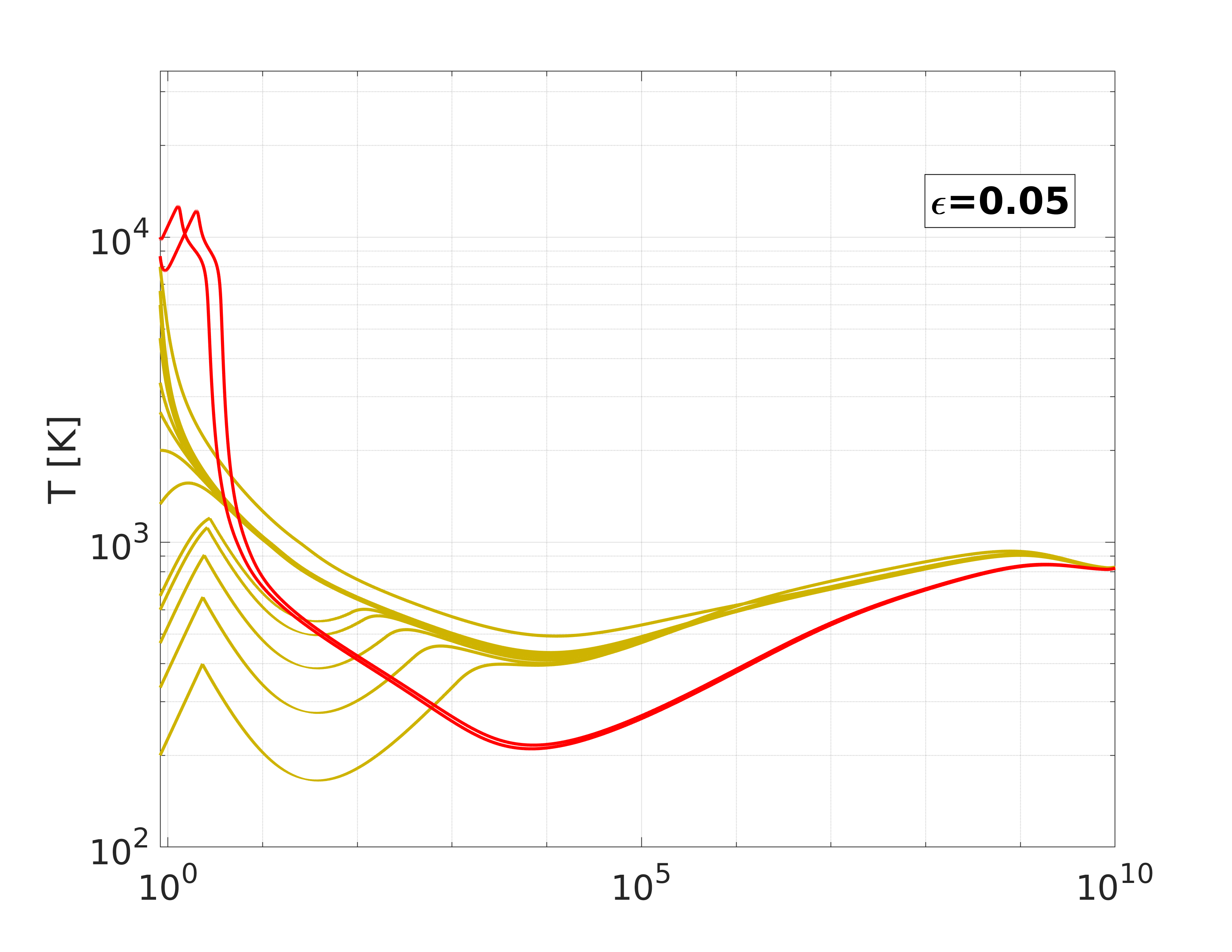} & 
			\includegraphics[width=0.47\textwidth]{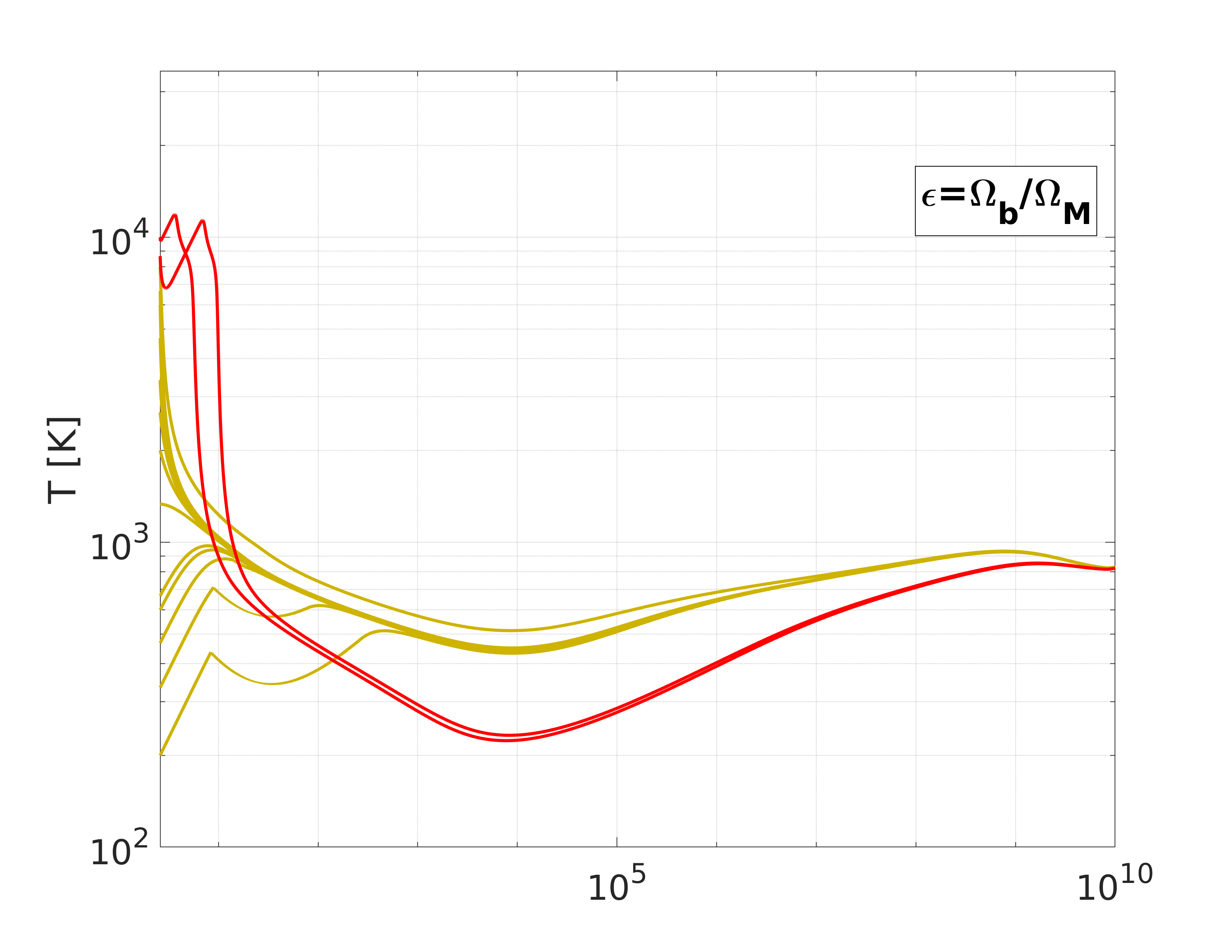} \\ 
			\includegraphics[width=0.47\textwidth]{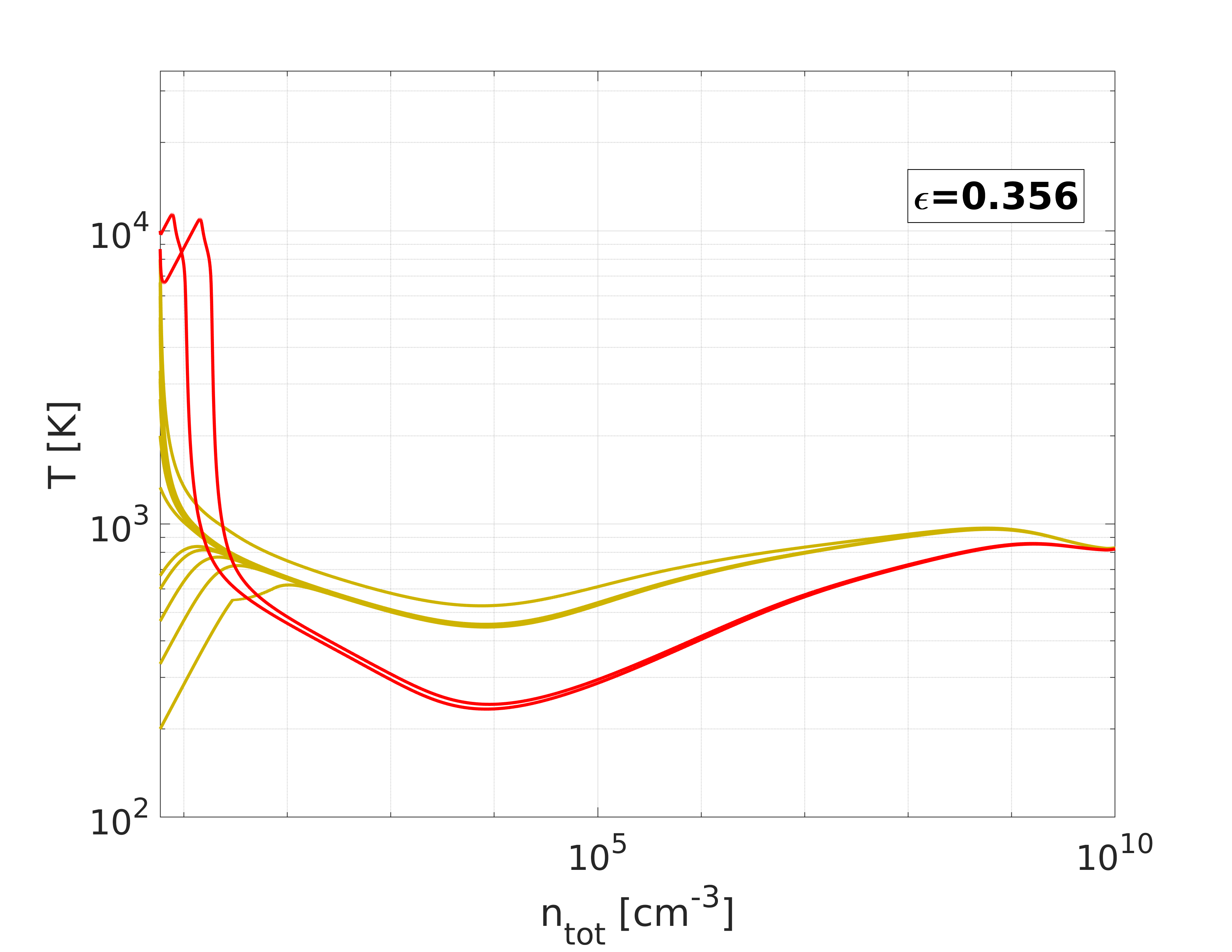} &
			\includegraphics[width=0.47\textwidth]{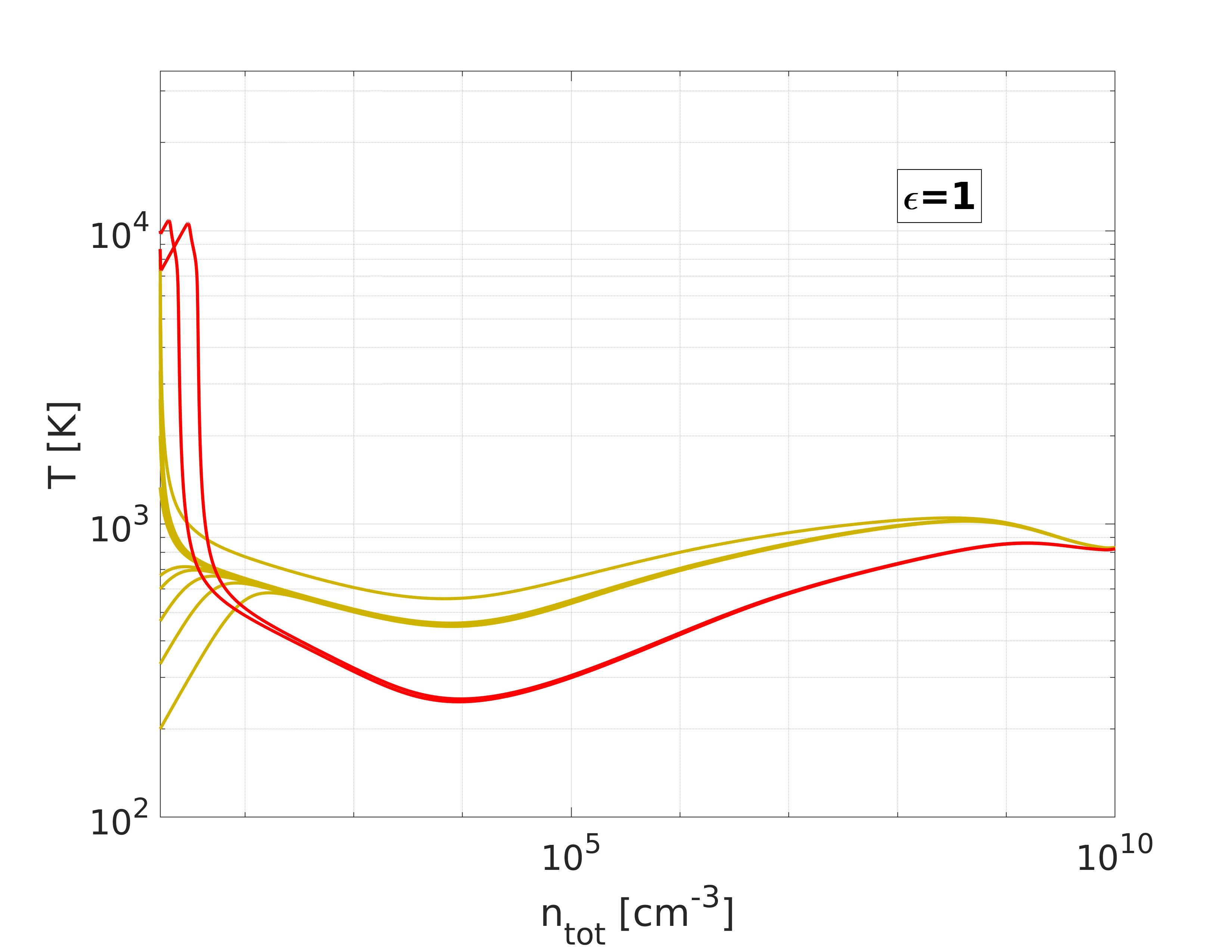} \\
		\end{tabular}
		
		\caption{
            The temperature and density evolution of the dissipative dark matter component in a one-zone collapse simulations, varying the dissipative dark matter fraction $\epsilon$, and using the primordial abundances from \citet{Gurian2021}. The simulations have otherwise identical initializations to the right column of Figure \ref{fig:damico_comp}. As in the previous figure, red lines denote efficient gas cooling, and yellow is quasi-isothermal collapse in the range $(500-900)\si{\kelvin}$. Significantly, no trajectories exhibit the high-temperature quasi-isothermal behaviour seen there, due to the greater initial $\hdt$ abundance. Varying $\epsilon$ most directly affects the initial total number density, with low temperature trajectories at lower $\epsilon$ values also demonstrating the alternating adiabatic/isobaric evolution that results in low-density cooling. The threshold value of $\epsilon=0.356$, below which the isobaric phase of cooling is significant, is derived from Eq.(\ref{eq:iso_condition}).}
		\label{fig:epsilon_comp}
	\end{figure*}

The initial $\hdt$ abundance has a critical role in the evolution of these halos. As mentioned, \cite{DAmico2017} used a low initial abundance of $x_{\hdt}=\num{e-10}$, assuming for simplicity that $x_{\ed} / x_{\hdt}=100$, independent of $\xi$ as in Standard Model cosmological recombination, and expecting that this was an overestimate of the true $\hdt$ fraction \citep{Latif2019}. By solving the background evolution equations, however, \cite{Gurian2021} have shown that, for Standard Model values of $m$, $M$, and $\alpha$ but $\xi=0.01$, the primordial $\hdt$ abundance is comparable to the Standard Model value, $x_{\hdt}\approx\num{e-6}$ at the time of structure formation, even though the free dark ion fraction is much lower. Thus, when using the primordial abundances from \cite{Gurian2021} we obtain the results in Figure \ref{fig:epsilon_comp}, where we have also varied $\epsilon$. The $\mathcal{O}(10^4)$ increase in $\hdt$ computed in \citet{Gurian2021}, over the value used in \cite{DAmico2017} and Figure \ref{fig:damico_comp}, ensures that halos with sufficient free electrons will undergo efficient molecular cooling before reaching the cooling-behavior density transition at $n_{\rm tot}\approx\SI{e5}{\pcc}$. That is, the trajectories that exhibited high-temperature quasi-isothermal behavior (blue) in Figure \ref{fig:damico_comp} are converted to trajectories with efficient cooling (red) in the top right panel of Figure \ref{fig:epsilon_comp}. 

The effect of varying $\epsilon$ is a bit more subtle. Primarily, the initial value of $n_{\rm tot}$ scales linearly with $\epsilon$, as does the total particle number density at recombination, $n_{\rm rec}$. The freeze-out free ion abundance is inversely related to $n_{\rm rec}$. With the other parameters (and hence $z_{\rm rec}$) fixed, this implies $x_e(\epsilon)\approx x_e(\epsilon=1) / \epsilon$. Meanwhile, $x_{\hdt}(\epsilon)\propto x_{\ed} n_{\rm form}$, where $n_{\rm form}$ is the number density at molecule formation. Since $z_{\rm rec}/z_{\rm form}$ is a constant, so is $n_{\rm rec}/n_{\rm form}$. Thus, $x_{\hdt}(\epsilon)\approx x_{\hdt}(\epsilon=1)$: decreasing $\epsilon$ increases the free ion abundance without changing the $\hdt$ abundance. 

Varying $\epsilon$ may also cause the halo to enter isobaric evolution, resulting in cooling. Essentially, the sound-crossing time behaves as $t_s\propto \left(\epsilon_M / \rho_{\rm DDM}\right)^{1/3} T^{-1/2}$, where $\rho_{\rm DDM}$ is the local dissipative dark matter density, $\epsilon_M=\epsilon\,\Omega_{DM} / \Omega_M$, while the free-fall time behaves as $t_{ff} \propto \left(\rho_M(1-\epsilon_M) + \rho_{\rm DDM}\right)^{-1/2}$, with $\rho_M$ the total local matter density. The densities are written separately here because $\rho_M (1-\epsilon_M)$ is constant in these simulations and in general may evolve differently than $\rho_{\rm DDM}$.  The isobaric transition condition can then be transformed into a temperature threshold $T_{iso}$, where
	\begin{equation}
		T_{iso} \propto \epsilon_M^{2/3} \frac{\rho_M(1-\epsilon_M)+\rho_{\rm DDM}}{\rho_{\rm DDM}^{2/3}},
		\label{eq:iso_condition}
	\end{equation}
above which isobaric evolution occurs. Thus, if the halo is heating adiabatically and crosses the temperature threshold, it generally cools until it can evolve adiabatically again. As the threshold temperature has an $\epsilon$ and $n$-dependent minimum, the threshold-crossing behavior produces a characteristic dip in the temperature at low densities, highly noticeable in the low temperature trajectories in the small-$\epsilon$ panels of Figure \ref{fig:epsilon_comp}. The $\epsilon=0.356$ panel demonstrates the maximum value of epsilon where this threshold-crossing behavior occurs, for the given initial conditions. Above this value, the trajectories never heat enough to cross the temperature threshold and so experience purely adiabatic evolution.

\begin{figure*}[htbp!]
		\centering
		\subfloat[$m=\SI{250}{\kilo\electronvolt}$, $M=\SI{20}{\giga\electronvolt}$, $\alpha=2/137$ and $\xi=0.02$]{%
            \includegraphics[width=0.47\textwidth]{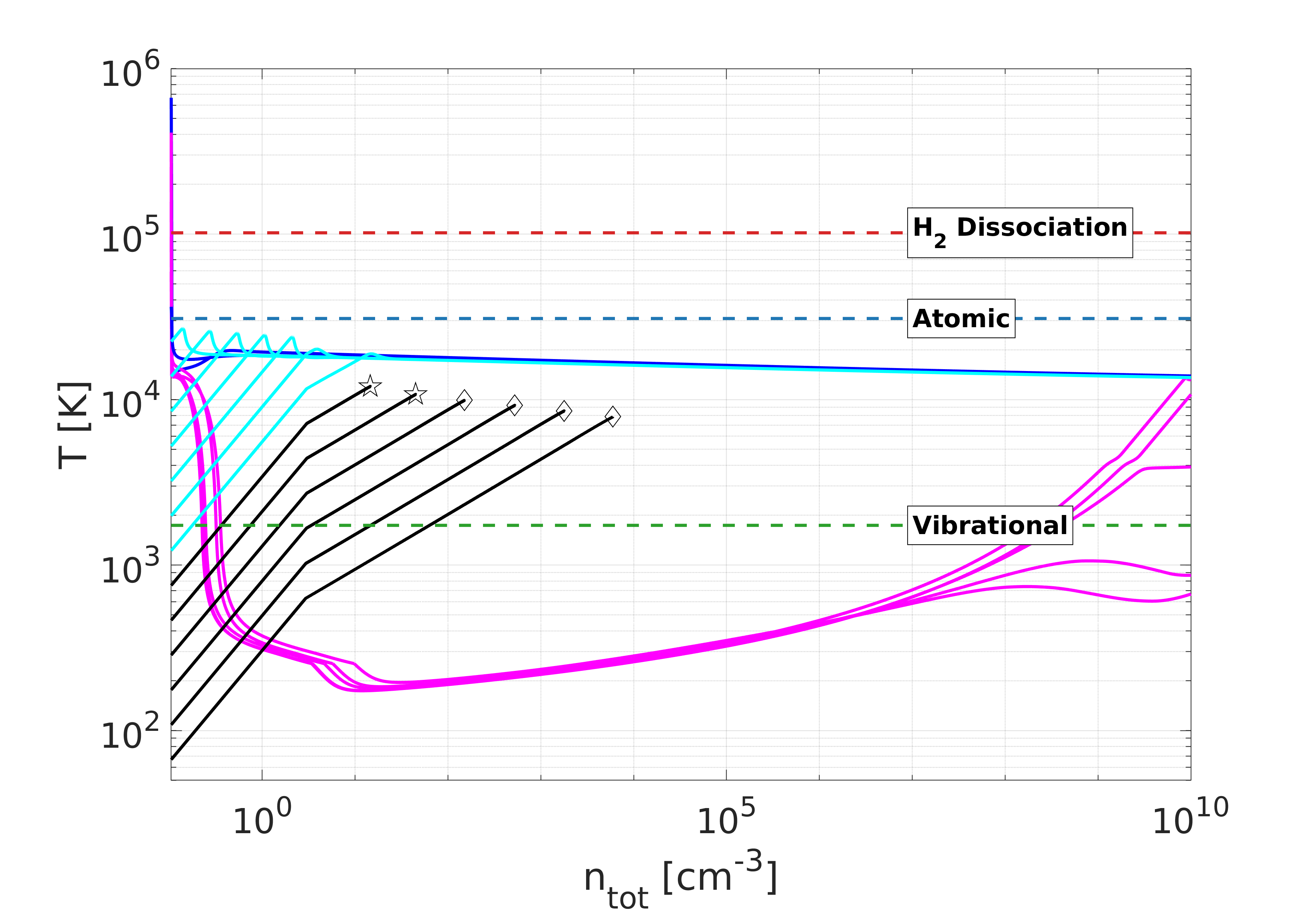}%
            \label{fig:dark_params1}%
        }
        \subfloat[$m=\SI{1}{\mega\electronvolt}$, $M=\SI{0.1}{\giga\electronvolt}$, $\alpha=137^{-1}$ and $\xi=0.05$]{%
            \includegraphics[width=0.47\textwidth]{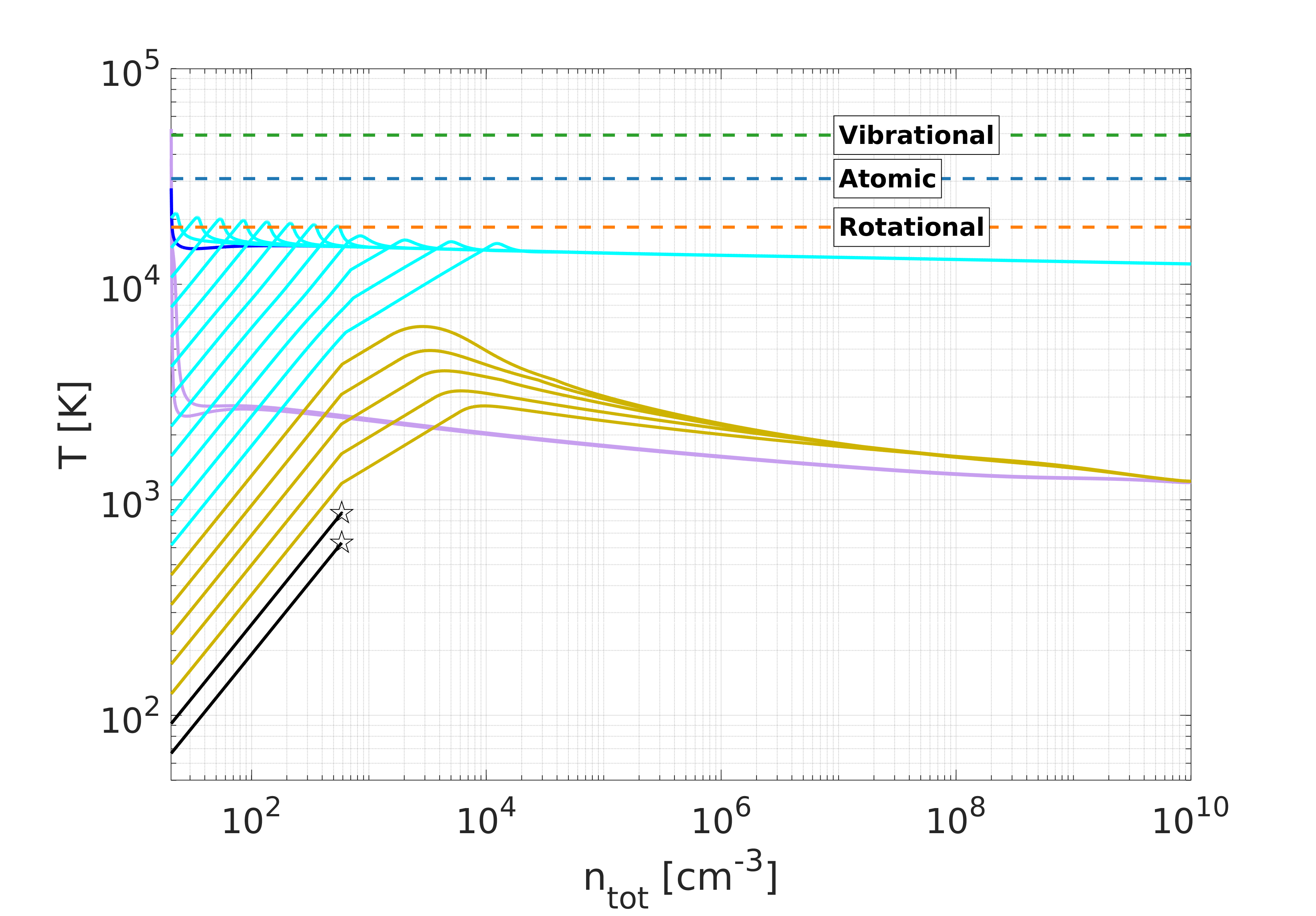}%
            \label{fig:dark_params2}%
        }
		
		\caption{
            Temperature versus total particle number density for the specified parameters. In the first panel, four behaviors are observed: pseudo-equilibrium (black with diamonds and stars), initial heating/cooling until atomic cooling balances compressional heating (cyan/blue), and efficient molecular cooling at low densities (magenta). In the second panel we see additional behaviors:  low-temperature quasi-isothermal evolution due to insufficient $\hdt$ production (mustard) and low-temperature quasi-isothermal cooling from rapid $\hdt$ production (lavender). Also plotted are various temperature thresholds as described in Figure \ref{fig:damico_comp}, re-scaled as described in \citet{Ryan2021}. In the second panel, since $T_{\rm vib}/T_{\rm C.E. peak} \approx 0.4 \sqrt{(\num{1836}\,m)/M} \approx 1.7$, the lowest vibrational energy transition occurs at a higher temperature than the atomic collisional-excitation-cooling peak.}
		
	\end{figure*}

	Lastly, in Figures \ref{fig:dark_params1}-\ref{fig:dark_params2} we consider dark matter that is fully dissipative ($\epsilon=1$) and demonstrate how varying the values of $m$, $M$, $\alpha$, and $\xi$ can drastically change the evolution of the halos. In Figure \ref{fig:dark_params1}, with $m=\SI{250}{\kilo\electronvolt}$, $M=\SI{20}{\giga\electronvolt}$, $\alpha=2/137$, and $\xi=0.02$ four different behaviors emerge. For these parameters, atomic collisional excitation cooling becomes efficient at approximately \SI{3000}{\kelvin}, so for halos with initial temperatures below that threshold (in cyan), the halo heats adiabatically until atomic cooling and compressional heating balance. Likewise, for the majority of temperatures above the threshold (in blue), the halo immediately cools until the balance is achieved. In both cases, insufficient $\hdt$ production prevents efficient molecular cooling. In a (relatively) small temperature range (in magenta), however, enough $\hdt$ is produced to cool the cloud, at least down to below the lowest vibrational transition. The halo is unable to cool down to the rotational regime before transitioning from low-density rovibrational cooling (which scales as $n^2$) to less-density-efficient high-density rovibrational cooling (which scales as $n$), around $n_{\rm tot}\approx\SI{1}{\pcc}$. Since compressional heating scales as $n^{3/2}$, as the density increases, it begins to dominate the thermal evolution. With minimal 3-body processes, and without tracking photons and $\text{H}_{D,3}$ reactions, the behavior above $n_{\rm tot}\approx\SI{e8}{\pcc}$, i.e. in the high density and opacity regime, is uncertain. Lastly, in the lowest temperature halos, the trace $\hdt$ production is sufficient for some initial cooling, seen when the trajectory switches from $n^{2/3}$ compressional heating to the adiabatic/isobaric oscillatory behavior (which follows $n^{1/3}$ behavior, as seen in Equation \ref{eq:iso_condition}). Unlike the low-temperature trajectories of Figure \ref{fig:epsilon_comp} and the cyan trajectories however, here the temperature is too low for atomic cooling to be relevant, and the molecular cooling has already entered the inefficient, high-density regime. The dissipative-dark-matter component of the halo thus enters a pseudo-equilibrium, where it is cooling back below the isobaric temperature threshold very inefficiently, taking longer and longer to re-enter the adiabatic phase. In this case, the halo remains in said state between \num{1} and $\SI{10}{\giga\year}$(black with diamond) or more than \SI{10}{\giga\year}(black with star), at time of simulation termination.
	
	In Figure \ref{fig:dark_params2}, with $m=\SI{1}{\mega\electronvolt}$, $M=\SI{0.1}{\giga\electronvolt}$, $\alpha=137^{-1}$, and $\xi=0.05$, we observe five categories of behavior, with only some of the behaviors in common with the previous parameter set. For starting temperatures below approximately \SI{6000}{\kelvin}, the halo collapses and heats adiabatically until $n_{\rm tot}\approx\SI{600}{\pcc}$, at which point halos begin alternating between adiabatic and isobaric evolution and the amount of generated $\hdt$ becomes critical. At the lowest temperatures, the halos enter pseudo-equilibrium, as before. However, for the starting temperatures in the approximately \SIrange{150}{800}{\kelvin} range, the halo does not form sufficient $\hdt$ early enough to prevent further heating and undergoes the low-temperature quasi-isothermal evolution seen in the Standard Model halos instead (mustard). Note that the later cooling from the long tail of the thermal population is occurring much further below the lowest rotational transition than in the Standard Model, as the rovibrational cooling channels (which re-scale inversely to the dark proton mass, see Equations \ref{eq:hdl_rot_scaling}-\ref{eq:ldl_scaling}) have significantly increased magnitude. At higher starting temperatures, halos achieve the high-temperature quasi-isothermal evolution seen in Figure \ref{fig:dark_params1}. At the highest starting temperatures, however, we observe the effect of coexisting atomic and molecular processes (lavender). Essentially, both the $\hd^-$ and $\hdt^+$ paths contribute to $\hdt$ formation, rapidly forming large amounts of $\hdt$, which, combined with the increased atomic cooling, drops the halo once again to the point of low-temperature quasi-isothermal evolution. 
	
\section{Conclusion and Outlook} \label{sec:conclusion}
	We have created an extension of the \krome \citep{Grassi2014} software package that enables the inclusion of dark sector chemistry in simulations. \dk aims to provide a drop-in replacement for \krome with expanded functionality that can flexibly add dark reactions and thermal processes while still solving the rate equations and providing additional \krome features like charge balancing and reaction checking. 
	
	We demonstrated that we can reproduce results found in the literature on \ddm simulations and can use \dk to explore the parameter space. We showed that the thermal evolution populations of a one-zone cloud collapse model are dependent on the initial abundances as well as the overall dissipative dark matter fraction, and differ from prior literature. Finally, we provided examples of other possible behavior populations that may be encountered at low particle densities in the dark parameter landscape.
	
	As the high-density dark chemistry  has not been fully determined at time of publication, we do not speculate on the end points of these clouds and leave more thorough cloud collapse simulations to future work. We anticipate \dk will find significant utilization in those simulations, and, due to the high extensibility of both it and KROME, will further assist other simulations involving dark chemistry, either in the atomic dark matter model or in other dissipative dark matter models.

	\acknowledgements
	Funding for this work was provided by the Charles E. Kaufman Foundation of the Pittsburgh Foundation. We thank Guido D'Amico for his input on how he used \krome in his simulations. We thank the anonymous referee for providing a truly excellent and professional report, which contributed significantly to the depth of analysis presented in the revised version.
	\software{KROME \citep{Grassi2014}}
	
	\appendix

\section{Atomic Cooling Process Rates} \label{sec:atomic_cooling}
	The \krome software package uses the chemical cooling and heating rates primarily found in \cite{Cen1992} for standard, baryonic matter. \dk provides two alternate sets of cooling rates for \ddmns: re-scaled versions of the Cen rates either directly from or based on the procedure in \citet{Ryan2021} (the \code{-cooling=DARKATOM} option), or the analytical expressions from \citet{Rosenberg2017} (the \code{-cooling=ADARKATOM} option). When $m=m_e=\SI{511}{\kilo\electronvolt}$, $M=m_p=\SI{0.938}{\giga\electronvolt}$, and $\alpha_D=\alpha=137^{-1}$, the rates very nearly agree. 
	
	In the following subsections, we discuss how all of the rates are re-scaled along with some implementation details for the recombination, collisional ionization, collisional excitation, and bremsstrahlung analytic rates. In the equations below, the cooling rate $\Lambda$ has units of \si{\epccm} and we assume chemical equilibrium in the figures. We use 
	\begin{align}
		\label{eq:ratios}
		\rc{}=\frac{m}{\SI{511}{\kilo\electronvolt}}\,,
		\rx{}=\frac{M}{\SI{0.938}{\giga\electronvolt}}\,,
		\ra{}=\frac{\alpha}{137^{-1}}\,.
	\end{align}
	
	In addition, we define a temperature re-scaled by the atomic energy scale as
	\begin{equation}
		\tilde{T}_a=\frac{T}{r_{\alpha}^2r_m}\,.
	\end{equation}
	
\begin{figure}[htbp]
		\centering
		\subfloat[]{%
            \includegraphics[width=0.32\textwidth]{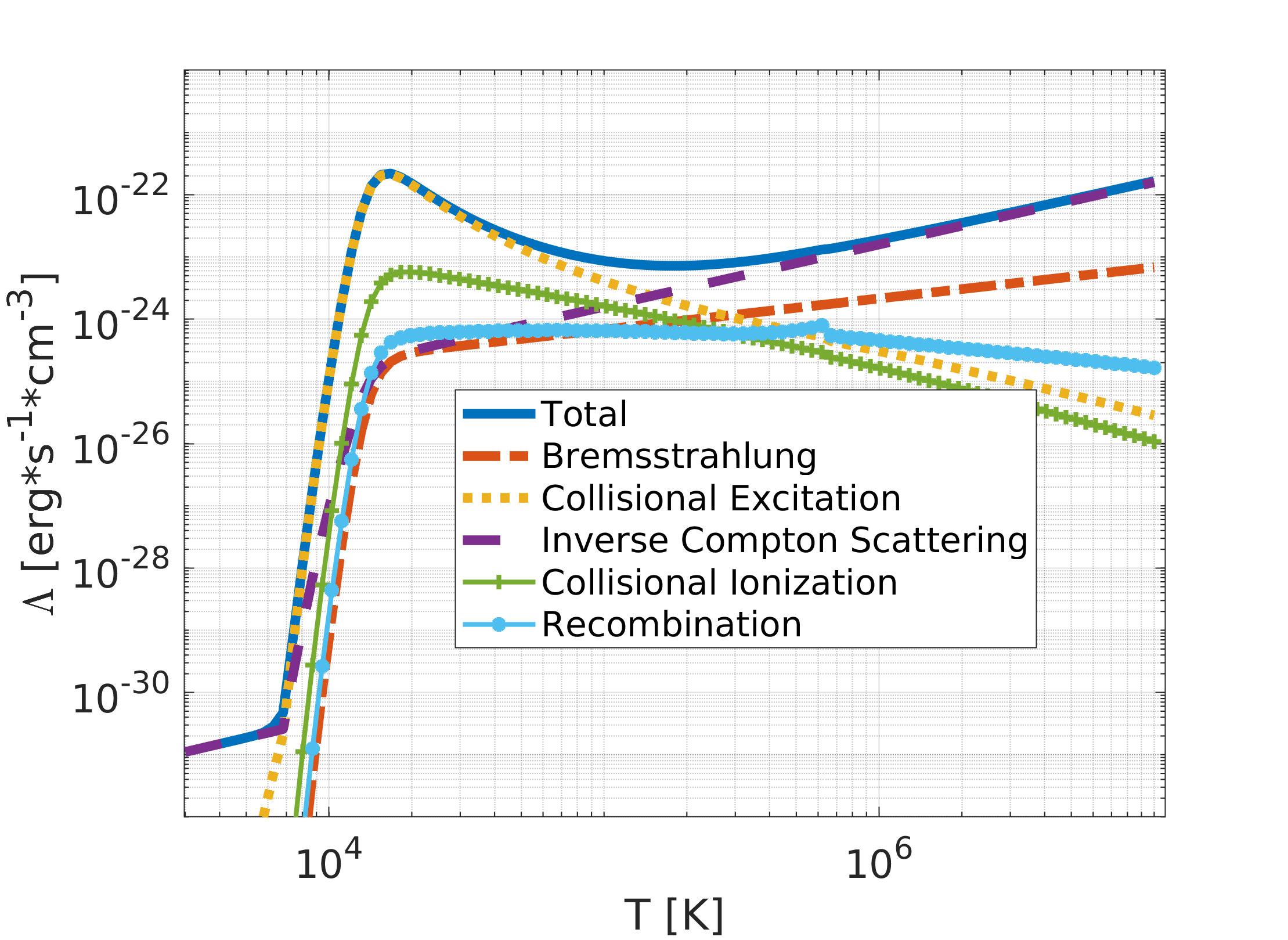}%
            \label{fig:DK_thermal_comp_sm}%
        }
        \subfloat[]{%
            \includegraphics[width=0.32\textwidth]{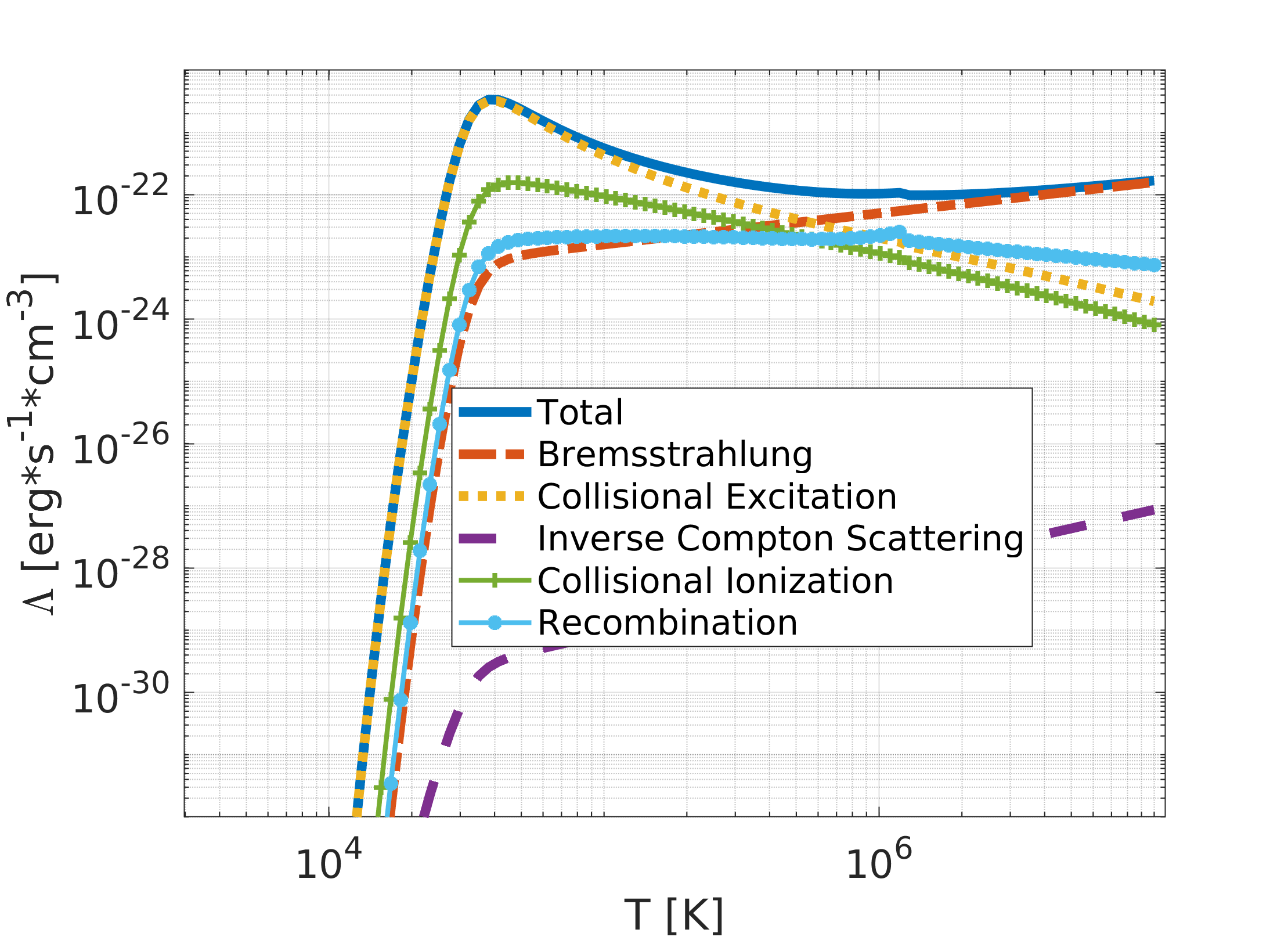}%
            \label{fig:DK_thermal_comp_dm1}%
        }
        \subfloat[]{%
            \includegraphics[width=0.32\textwidth]{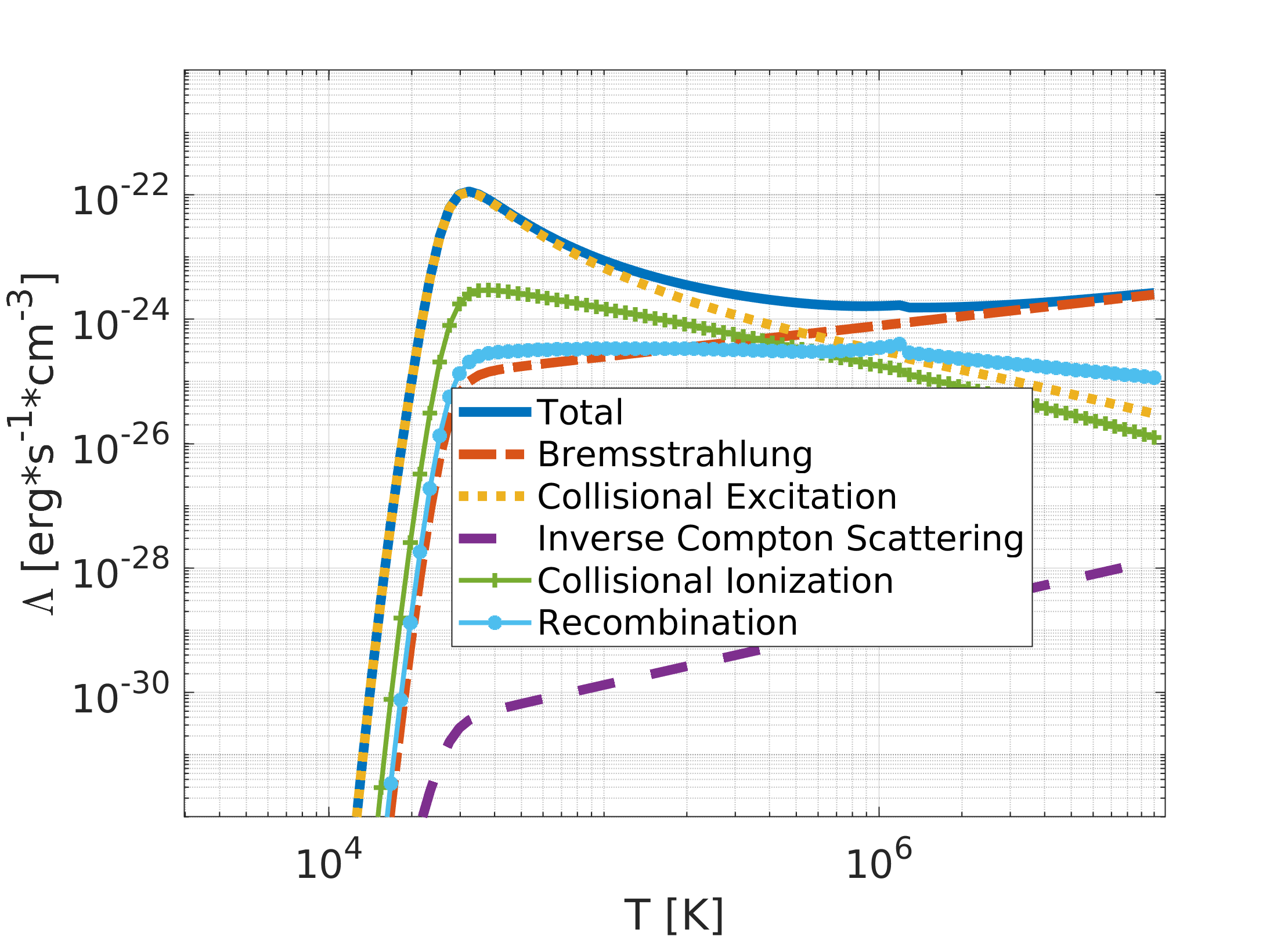}%
            \label{fig:DK_thermal_comp_dm2}%
        }
		\caption{Plots demonstrating the contributions to the net dark atomic cooling process and their parameter dependence. The parameter values are $(m /\si{\kilo\electronvolt},M/\si{\giga\electronvolt},\alpha,\xi)=\{(511,0.938,137^{-1},1),(250,20,2/137,0.02),(10^3,0.1,137^{-1},0.05)\}$, corresponding to panels \{\ref{fig:DK_thermal_comp_sm},\ref{fig:DK_thermal_comp_dm1},\ref{fig:DK_thermal_comp_dm2}\} and the parameter values used in Figures \{\ref{fig:onezonekromevdk},\ref{fig:dark_params1},\ref{fig:dark_params2}\}. In all cases, the dark gas is assumed to be in chemical equilibrium, the total particle density is $n_{\rm tot}=\SI{1}{\pcc}$, and redshift is $z=40$. The slight kink in the recombination rate results from the low-to-high temperature-limit transition. Note the increased cooling in panel (b), the temperature shifts from the change in the dark hydrogen binding energy, and the strong $\xi$ dependence of inverse Compton scattering.}
		\label{fig:DarkKROME_thermal_process_comparison}
	\end{figure}

	In Figure \ref{fig:DarkKROME_thermal_process_comparison}, we demonstrate how the individual components contribute to the overall cooling rate, $\Lambda$, for Standard Model values and the two sets of dark parameters used in Section \ref{sec:results}, $(m,M,\alpha,\xi)=\{(\SI{250}{\kilo\electronvolt},\SI{20}{\giga\electronvolt},2/137,0.02),\,(\SI{1}{\mega\electronvolt},\SI{0.1}{\giga\electronvolt},137^{-1},0.05)
	\}$.
	
\subsection{Recombination} \label{sec:recomb}
	The dark atomic recombination ($\hd^++\ed\rightarrow \hd+\gamma_D$) cooling rate is simply given as the thermal average of the collision kinetic energy summed over all energy levels \citep{Rosenberg2017}, or 
	\begin{equation}
		\Lambda_{\rm rec} = \sum_{n=1}^{\infty}\langle (\KE)\, \sigma_{{\rm rec},n} v \rangle n_{\hd^+}n_{\ed}.
	\end{equation}
	To compute the re-scaled rate then, we need the $\sigma_{{\rm rec},n}$ re-scaling, which, from \citet{Ryan2021}, is 
	\begin{equation}
		\sigma_{{\rm rec},n,{\rm DM}} = \ra{5}\rde^{-2}\sigma_{{\rm rec},n}.
	\end{equation}
	The kinetic energy term introduces an additional factor of $T$, so the final re-scaling is 
	\begin{align}
		\Lambda_{\rm rec, DM}(T) &= \sqrt{\frac{\rde}{\rc{}}} \rde \left(\ra{5}\rde^{-2}\right) \Lambda_{\rm rec}\left(\frac{T}{\rde}\right) \\
		&= \rac{5}{-1/2} \rde^{-1/2} \Lambda_{\rm rec}\left(\frac{T}{\rde}\right) \\
		&= \rac{4}{-1} \Lambda_{\rm rec}(\tilde{T}_a),
		\label{eq:recomb_rescale}
	\end{align}
	where the first term in the first line comes from the $\sqrt{T/m}$ term in the thermal average and the second from the collision kinetic energy, and in the third line we used $\inc E \propto m\alpha^2$. For the analytic rate, we avoid the full integral calculation from \citet{Rosenberg2017} to decrease computation time and simply use their high and low limits (defining $y^2=(m\, \alpha^2)/(2 k_B T)$, where $k_B$ is the Boltzmann constant),
	\begin{equation}
		\frac{\Lambda_{\rm rec}}{n_{\ed} n_{\hd^+}} = \begin{cases}
			4.7\times 10^{-25}\,\rac{3}{-3/2} \left(\frac{T}{10^5}\right)^{1/2} (0.74 + \log{y^2} +\frac{1}{3 y^2}) & y\gg 1\\
			1.1\times 10^{-25} \,\rac{5}{-1/2} \left(\frac{10^6}{T}\right)^{1/2} (5+y^2 (2.860+14/3 \log{y^2})) & y\ll 1 
		\end{cases}.
		\label{eq:recomb}
	\end{equation}
	Transitioning at $y^2=1/4$ provides a good fit to the full integral, varying at most by approximately 40 percent at the transition point. 

\subsection{Collisional Ionization} \label{sec:coll_ion}
\begin{figure}[htbp]
		\centering
		\includegraphics[width=0.5\linewidth]{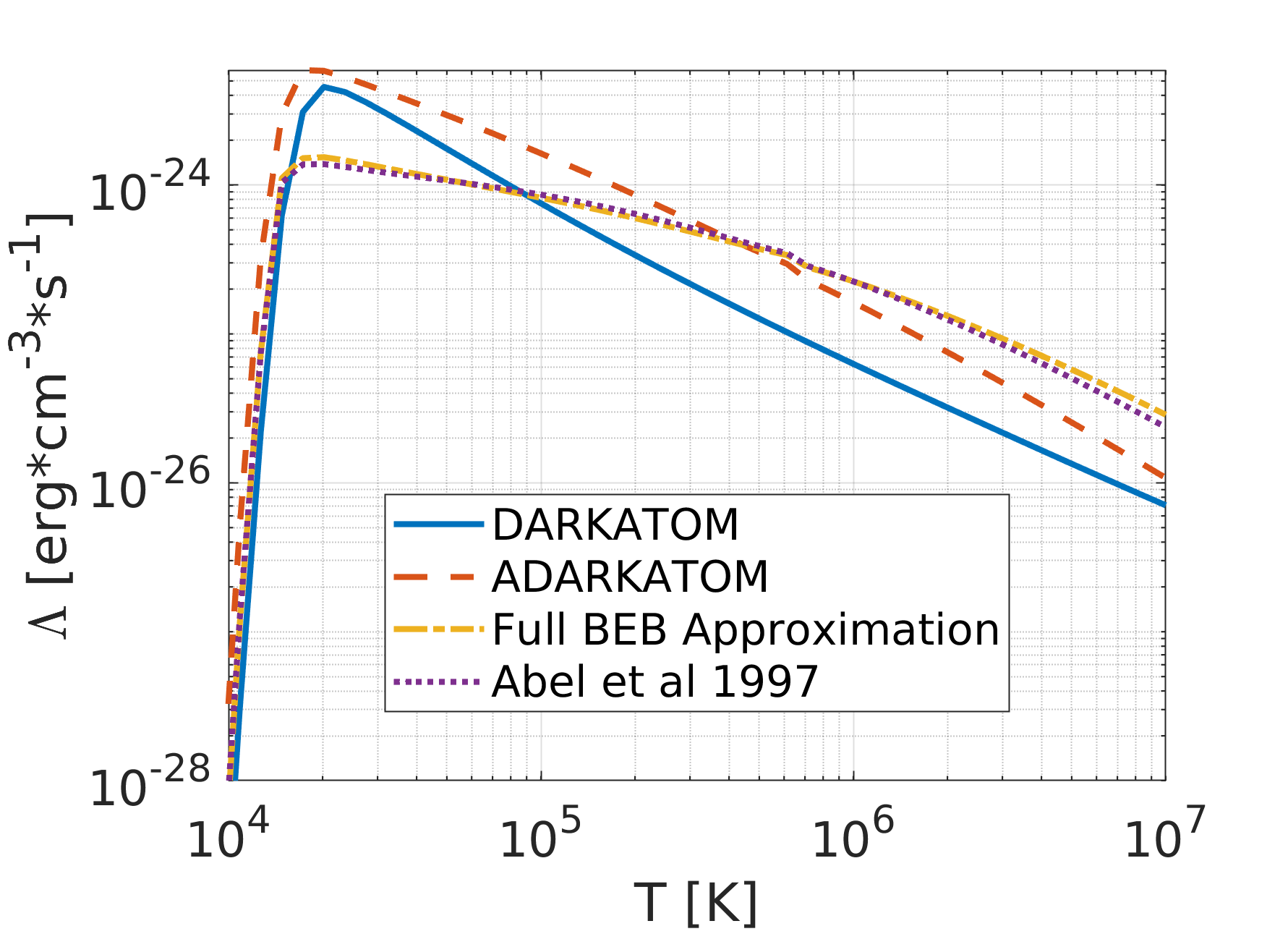}
		\caption{Comparison of the \code{DARKATOM}, \code{ADARKATOM}, and binary-encounter-Bethe collisional ionization rates for $m=\SI{511}{\kilo\electronvolt}$ and $\alpha=137^{-1}$. The re-scaled \krome rate in \code{DARKATOM} (solid blue) is based on \citet{Cen1992}, while the analytic rates (red dashed and yellow dot-dashed) are from \cite{Rosenberg2017}. We also plot a more recent rate from \citet{Abel1997} (dotted purple), demonstrating that the differences between the various dark rates are comparable to the difference between the \citet{Cen1992} and \citet{Abel1997} rates.}
		\label{fig:KROMEvsDARKcollion}
	\end{figure}

	Like the recombination cooling rate, the collisional ionization cooling rate is simply the reaction rate multiplied by the energy lost, $\inc E \propto m\alpha^2$. To compute the re-scaled reaction rate, we use the simple binary encounter approximation \citep{Peterkops1977} to obtain the overall parametric dependence of the cross section, followed by the re-scaling procedure. This gives a re-scaled cross section of $\sigma_{\rm ci,D} = \rac{-2}{-2} \sigma_{\rm ci}$ and the final re-scaling is then 
	\begin{align}
		\Lambda_{\rm ci, DM}(T) &= \sqrt{\frac{\rde}{\rc{}}}\rde \left(\rac{-2}{-2}\right)\Lambda_{\rm ci}\left(\frac{T}{\rde}\right) \\
		&= \rac{}{-1} \Lambda_{\rm ci}(\tilde{T}_a).
	\end{align}
	While \citet{Rosenberg2017} also provides the more accurate binary-encounter-Bethe model for the analytic rate, we continue to use the binary encounter approximation, which has an analytic solution involving exponential integrals (${\rm Ei}(z)=-\int_{-z}^{\infty } e^{-t} / t \, dt$),
	\begin{align}
		\frac{\Lambda_{\rm ci}}{n_{\ed} n_{\hd}} &=  \num{3.9e-18} \rac{2}{-1/2}\sqrt{\frac{\SI{e5}{\kelvin}}{T}}f(y^2) \\
		f(y^2) &= \frac{1}{2}\left(e^{-y^2} + y^2 {\rm Ei}(-y^2)\right).
	\end{align}
	As a well-known special function, ${\rm Ei}(z)$ can be rapidly calculated from library functions, avoiding the computational slow-down of the more complicated binary-encounter-Bethe model, with only an $\mathcal{O}(1)$ difference in magnitude \citep{Rosenberg2017}, comparable to the difference between different Standard Model rate sources. In Figure \ref{fig:KROMEvsDARKcollion} we compare the re-scaled \citet{Cen1992} rate used in the \code{-cooling=DARKATOM} option, the analytic rate based on the binary encounter approximation used in the \code{-cooling=ADARKATOM} option, the full binary-encounter-Bethe model, and a more recent rate from \citet{Abel1997} for Standard Model values of $m$ and $\alpha$. 
	
\subsection{Collisional Excitation} \label{sec:coll_exc}
	To compute the re-scaled collisional excitation cooling rate, we use the same overall cross-sectional re-scaling as collisional ionization, $\sigma_{\rm ce, D} = \rac{-2}{-2} \sigma_{\rm ce}$. This matches the dependence found using the Born approximation combined with empirical scaling \citep{Schiff1968,Kim2001,Rosenberg2017}, and leads to the same overall rate re-scaling,
	\begin{align}
		\Lambda_{\rm ce, DM}(T) &= \rac{}{-1} \Lambda_{\rm ce}(\tilde{T}_a).
	\end{align}
	Analytically, the $1s\rightarrow 2p$ transition dominates over all other ground state transitions and so we ignore other transitions and keep only the leading order term when computing the final rate (with $y^2=(m \alpha^2)/(2 k_B T)$),
	\begin{align}
		\label{eq:gfunc}
		\frac{\Lambda_{\rm ce}}{n_{\hd^+} n_{\hd}} &= \num{3.9e-18}\,\rac{2}{-1/2} \sqrt{\frac{10^5}{T}} g(y^2)  \\
		g(y^2) &= \int_{\frac{\sqrt{3}}{2} y}^{\infty} \text{d} u \frac{u \text{e}^{-u^2}}{1+\frac{7 y^2}{4 u^2}} \log{\frac{4 u}{y}}
		\label{eq:coll_exc}
	\end{align}
	Note that this rate has a higher magnitude than the \citet{Cen1992} rate when calculated using Standard Model values, but as demonstrated by \citet{Rosenberg2017}, the analytic formula has the same level of agreement with the Cen rate as a newer Standard Model rate from \citet{Callaway1994}.

\subsection{Bremsstrahlung} \label{sec:bremss}
	The nonrelativistic, thermal bremsstrahlung cooling rate, up to a Gaunt factor, has an analytic expression of the form \citep{Rybicki1985},
	\begin{align}
		\Lambda_{\rm ff,D}(T) &= \sqrt{\frac{2\pi k_B T}{3 m}} \frac{16 \hbar^2 \alpha^3 c}{3 m} n_{\ed} n_{\hd^+} g_{ff,D} \\
		&= \num{1.4e-27}\,\rac{3}{-3/2} \left(\frac{T}{1 \text{K}}\right)^{1/2} n_{\ed} n_{\hd^+} g_{ff,D}\\
		&= \rac{4}{-1} \Lambda_{\rm ff}(\tilde{T}_a).
		\label{eq:brems}
	\end{align}
	That is, the analytic expression from \citet{Rosenberg2017} and the re-scaled rate are identical. The default value of $g_{ff,D}=\num{1.5}$ (\code{darkGauntFF}) matches the value used in \krome.

\subsection{Inverse Compton Scattering} \label{sec:compton}
	Like bremsstrahlung, in the nonrelativistic, low-energy limit, the cooling rate due to inverse Compton scattering has an analytic form \citep{Mo2010},
	\begin{align}
		\Lambda_{\rm ics,D} &= \frac{4 k_B (T_{\ed}-T_{\gamma_D})}{m c^2} c \left[\frac{8\pi}{3} \left(\frac{\alpha \hbar}{m c}\right)^2\right] n_{\ed} a_r T_{\gamma_D}^4 \\
		&= \num{1.0e-37}\left(\frac{T-T_{\gamma_D}}{\SI{1}{\kelvin}}\right) \rac{2}{-3} \left(\frac{T_{\gamma_D}}{\SI{1}{\kelvin}}\right)^4 \\
		&= \rac{2}{-3}\Lambda_{\rm ics}(T,T_{\gamma_D}),
	\end{align}
	so the re-scaled rate is identical to the rate given in \citet{Rosenberg2017}. We assume the dark photon temperature is set to background, i.e. $T_{\gamma_D}=(1+z)T_{\gamma,D} = (1+z)\,\xi\, T_{\rm CMB}$ for redshift $z$.

\section{Dark Molecular Processes} \label{sec:dmc} 
	For molecular processes \dk uses the rates derived in \cite{Ryan2021}. The added molecular reactions are listed in Table \ref{tab:reacts_dark} and we describe our implementation of the thermal processes here. We also give a brief derivation of the $H_2 + H\rightarrow 3H$ reaction scaling. Both $H_2$ collisional cooling and the chemical reaction cooling processes are included by selecting \code{-cooling=DARKMOL}.

\subsection{$H_2$ Collisional Cooling} \label{sec:h2_col_cool}
	$H_2$ collisional cooling, also referred to as line cooling, involves the collisional excitation of $H_2$ and subsequent radiative cooling. Our basic approach to implementing the dark version involves re-scaling the line cooling rates for $\hdt-\{\hd,\hdt,\ed,\hd^+\}$ collisions from \cite{Glover2015} using the process described in \cite{Ryan2021}. Effectively, this can be condensed to re-scaling the high-density rotational and vibrational cooling rates, $H_{R}, H_{V}$ (using the \krome notation \cite{Grassi2014})
	\begin{align}
		H_{R,D}(T) &= \racx{9}{8}{-6} H_{R}(\tilde{T}_{R}) \label{eq:hdl_rot_scaling}\\
		H_{V,D}(T) &= \racx{9}{5}{-3}H_{V}(\tilde{T}_{V}),
		\label{eq:hdl_vib_scaling}
	\end{align}
	where $\tilde{T}_{R}=\racx{2}{2}{-1}\,T$ and $\tilde{T}_{V}=\racx{2}{3/2}{-1/2}\,T$,  the low-density rovibrational cooling rate, $\Lambda_{\hdt,\{\hd,\hdt\}}$ as 
	\begin{equation}
		\Lambda_{\hdt,\{\hd,\hdt\}} = \begin{cases}
			\racx{}{}{-2}\; \begin{cases}
				\Lambda_{\text{H}_{2},\{\text{H},\text{H}_{2}\}}\left(\tilde{T}_r\right)\; & T \le T_{0,r} \\
				\text{lerp}\left(\Lambda_{\text{H}_{2},\{\text{H},\text{H}_{2}\}}\left(\tilde{T}_r	\right),T_{0,r},\tilde{T}_r\right) & T_{0,r} \le T \le T_0 
			\end{cases}\\
			\racx{}{1/4}{-5/4}\;\begin{cases}
				\text{lerp}\left(\Lambda_{\text{H}_{2},\{\text{H},\text{H}_{2}\}}\left(\tilde{T}_v	\right),T_{0,v},\tilde{T}_v\right)& T_0 < T \le T_{0,v} \\
				\Lambda_{\text{H}_{2},\{\text{H},\text{H}_{2}\}}\left(\tilde{T}_v\right) & T_{0,v} \le T
			\end{cases}
		\end{cases},
		\label{eq:ldl_scaling}
	\end{equation}
	the $\Lambda_{\hdt,\ed}$ cooling rate as
	\begin{equation}
	    \Lambda_{\hdt,\ed} = \rax{1}{-1}\Lambda_{\text{H}_2,\text{e}}(\tilde{T}_r),
	    \label{eq:ldl_e_scaling}
	\end{equation}
	and the $\Lambda_{\hdt,\hd^+}$ cooling rate as
	\begin{equation}
	    \Lambda_{\hdt,\hd^+} = \racx{1}{1/2}{-3/2}\Lambda_{\text{H}_2,\text{H}^+}(\tilde{T}_r),
	    \label{eq:ldl_p_scaling}
	\end{equation}
	Here we have defined $T_0$ as \SI{855.833}{\kelvin} for $H_2-H$ and \SI{5402.44}{\kelvin} for $H_2-H_2$, $T_{0,r}=\racx{2}{2}{-1}\,T_0$, $T_{0,v}=\racx{2}{3/2}{-1/2}\,T_0$, and $\text{lerp}(f(x),x_0,x)$ as the linear extrapolation of the function $f(x)$ from the point $x_0$. This is the dark equivalent to the cooling found in \code{-cooling=H2}.
	
	The net low-density rovibrational cooling rate consists of the sum of the $\hdt-{X_i}$ collisional rates with various species $\{X_i\}$. Of note, the $H_2$ rovibrational cooling from \cite{Glover2015} used in \krome also includes the cooling terms from collisions with helium, assuming it is present in the simulation. Since our dark matter model does not contain neutrons, and thus no helium, this cooling channel has been omitted.

\subsection{Endo- and Exoergic Processes} \label{sec:chem_cool}
	As described in \citet{Omukai2000,Grassi2014}, certain endo- and exoergic reactions contribute significantly to the thermal evolution of the gas. Some of these reactions are included in other heating and cooling options, like atomic collisional ionization and recombination, but the remainder are considered the ``chemical" cooling and heating processes and listed in Table \ref{tab:chem_thermo}.  
	
	\begin{table}[htbp]
		\centering
		\begin{threeparttable}
			\centering
			\begin{tabular*}{\columnwidth}{l r c c l r c}
				\toprule
				\textbf{Reaction} & \textbf{Energy/\si{\electronvolt}} & \textbf{Notes} & \hspace{2 cm} & \textbf{Reaction} & \textbf{Energy/\si{\electronvolt}} & \textbf{Notes}\\
				\midrule
				$\hdt + \hd \rightleftharpoons 3 \hd$ & \num{-4.48} & \tnote{1} & & $\hd^- + H \rightarrow H_2 + \ed$ & \num{3.53} &\\
				$\hdt + \ed \rightarrow 2\hd + \ed$ & \num{-4.48} & \tnote{2} & & $\hdt^+ + \hd \rightarrow \hdt + \hd^+$ & \num{1.83} & \\
				$2 \hdt \rightleftharpoons \hdt + 2\hd$ & \num{-4.48} & \tnote{1}\tnote{,} \tnote{3} & & & \\
				\bottomrule
			\end{tabular*}
			\caption{Reactions considered part of the ``chemical" heating and cooling processes. Reactions with negative energies cool the gas, positive energies heat. From \citet{Ryan2021}, all reactions have energy re-scalings given by $\ed=\rac{2}{}E_{SM}$.}
			\label{tab:chem_thermo}
			\begin{tablenotes}[flushleft]
				\item [1] Forward reaction provides cooling, inverse reaction provides heating.
				\item [2] Reaction not included in minimal reaction network.
				\item [3] Forward reaction not included in minimal reaction network.
			\end{tablenotes}
		\end{threeparttable}
	\end{table}
	
	The basic rate equation is quite simple, with 
	\begin{equation}
		\Lambda_j = E_j \, k_j n(R_{j1}) n(R_{j2}),
	\end{equation}
	where $E_j$ the energy consumed/produced and $k_j$ is the rate for reaction $j$, and the net rate is just $\Lambda_{\rm chem}=\sum_j \Lambda_j$. The dark analogs of $E_j$ and $k_j$ have already been computed in \citet{Ryan2021}. The rates in \krome are weighted by a critical density factor, $f=\left(1+n_{\rm cr}/n_{\rm tot}\right)$, however, following \citet{Hollenbach1979}, whose parametric dependence must still be determined. The critical density is approximated in the Standard Model rate as 
	\begin{equation}
		n_{\rm cr} = \frac{A_{\rm vib}}{\gamma_{20}^{\rm H} x_{\rm H} + \gamma_{10}^{\rm H_2} x_2},
	\end{equation}  
	with $A_{\rm vib}$ the Einstein $A$ coefficient, $x_{\rm H},x_2$ the relative abundances of $\rm H$ and $\rm H_2$, and $\gamma_{\inc v}^{\rm H,H_2}$ the collisional de-excitation rate coefficients for the $\inc v$ vibrational transition. From \cite{Ryan2021}, these quantities all re-scale with the dark parameters with the re-scalings described therein, and so the critical density re-scales as 
	\begin{equation}
	\begin{aligned}
	    n_{\rm cr,D} &= \racx{8}{19/4}{-7/4}n_{\rm cr}(\tilde{T}_v) \label{eq:crit_n_scaling}\\
		\tilde{T}_v &= \racx{-2}{-3/2}{1/2}\,T.
	\end{aligned}
	\end{equation} 
	Since the network used here from \cite{Ryan2021} only contains a minimal set of reactions, only the reactions $\hd^- + \hd \rightarrow \hdt + \ed$, $\hdt^+ + \hd \rightarrow \hdt + \hd^+$, $\hdt+2\hd\rightarrow2\hdt$, and $\hdt+\hd \rightleftharpoons 3 \hd$ contribute to the cooling and heating in the simulations presented here. The cooling and heating reactions are included if either \code{-cooling=DARKMOL} or \code{-heating=DARKMOL} option is specified, as the dark analog of the equivalent options \code{-cooling=CHEM} or \code{-heating=CHEM}.

\subsection{$H_2$ Dissociation Scaling} \label{sec:h2_diss}
	From \citet{Ryan2021}, the overall rate re-scaling for the collisional reaction $\text{H}_2 + \text{H} \rightarrow 3 \text{H}$ is given by 
	\begin{equation}
	    k_{\rm diss,D}(T) = \racx{-1}{-3/2}{-1/2}\; k_{\rm diss}(\tilde{T}_a),
		\label{eq:h2_diss_scaling}
	\end{equation}
	
	The Standard Model rate used in both \dk and \krome is based on a master rate formulation from \citet{Martin1996}, with the general form
	\begin{equation}
	    \log(k_{\rm diss}(T)) = \log(k_{h}(T))-\frac{\log(k_{h}(T))-\log(k_{l}(T))}{1+(n_H/n_{\rm cr}(T))^p},
	\end{equation}
	where $k_{h,l}(T)$ are the high(low)-density limits, $n_H$ is the $\rm H$ number density, and $n_{\rm cr}$ is the critical density at which downwards energy transitions switch from radiative to collisional. Thus, for the dark rate, we also need to re-scale the critical density using both the overall and temperature re-scaling from Equation \ref{eq:crit_n_scaling}.

\section{Reactions file} \label{sec:reactions} 
	\dk includes a basic reactions file, \code{react\_dark} containing several example reactions and parameter definitions. The included reactions are listed in Table \ref{tab:reacts_dark}, while the parameter definitions include setting $\xi=0.01$ and the dark free-free gaunt factor $g_{ff,D}=1.5$, used in dark bremsstrahlung. The format of the actual reactions is identical to that of \kromens, using the $Q$ notation as described in Section \ref{sec:krome_and_dk}.  
	
	\begin{table}[htbp!]
		\centering
		\begin{threeparttable}
			\begin{tabular*}{\textwidth}{c l c c c l c}
				\toprule
				& \textbf{Reaction} & \textbf{Source} & \hspace{1cm} &  & \textbf{Reaction} & \textbf{Source} \\
				\midrule
				1 & $\hd^++ \ed\rightarrow \hd + \gamma_D$ & \cite{Rosenberg2017} & & 10 & $\hdt^+ + \hd \rightarrow \hdt+\hd^+$ &\cite{Ryan2021}\\
				2 & $\hd + \gamma_D \rightarrow \hd^++\ed$ & \cite{Rosenberg2017} & & 15 & $\hdt + \hd^+\rightarrow \hdt^+ + \hd$&\cite{Ryan2021} \\
				\tnote{1} & $\hd+\ed\rightarrow \hd^++ 2\ed$ & \cite{Rosenberg2017} &  & 18 & $\hdt + \gamma_D \rightarrow \hdt^+ + \ed$ & \cite{Ryan2021} \\
				3 & $\hd + \ed\rightarrow \hd^- + \gamma_D$ &\cite{Ryan2021} & & \tnote{1} & $\hdt + \hd \rightarrow 3 \hd $ & \citet{Ryan2021}\tnote{2}\\
				4 & $\hd^- + \gamma_D \rightarrow \hd + \ed$ &\cite{Ryan2021} & & \tnote{1} &$3 \hd \rightarrow \hdt + \hd $  & \citet{Ryan2021}\\
				5 & $\hd^- + \hd \rightarrow \hdt + \ed$&\cite{Ryan2021} & & \tnote{1} & $\hdt + 2\hd \rightarrow 2 \hdt $ & \citet{Ryan2021}\\
				7 & $\hd^- + \hd^+\rightarrow 2 \hd$ &\cite{Ryan2021} & & \tnote{1} &$2\hd + \hd^+ \rightarrow \hdt + \hd^+ $  & \citet{Ryan2021}\\
				8 & $\hd + \hd^+\rightarrow \hdt^+ + \gamma_D$ &\cite{Ryan2021} & & \tnote{1} & $2\hd + \hd^+ \rightarrow \hdt^+ + \hd $ & \citet{Ryan2021} \\
				9 & $\hdt^+ + \gamma_D \rightarrow \hd + \hd^+$ & \cite{Ryan2021} & & & &\\
				\bottomrule
			\end{tabular*}
			\caption{Dark particle reactions included in the \dk \code{react\_dark} file. Reactions are numbered here according to \cite{Galli1998}.}
			\label{tab:reacts_dark}
			\begin{tablenotes}
				\item[1] Not included in \cite{Galli1998}.
				\item[2] See also Appendix \ref{sec:h2_diss}.
			\end{tablenotes}
			
		\end{threeparttable}
	\end{table}
	
\bibliographystyle{aasjournal}
\bibliography{darkkrome.bib}

\end{document}